\documentclass[twocolumn]{aastex631}

\shorttitle{Spectroscopic Orbits of Subsystems. IX}

\begin{document}

\renewcommand{\topfraction}{1.0}
\renewcommand{\bottomfraction}{1.0}
\renewcommand{\textfraction}{0.0}

\newcommand{\kms}{km~s$^{-1}$\,}
\newcommand{\masyr}{mas~s$^{-1}$\,}
\newcommand{\msun}{$M_\odot$\,}

\title{Spectroscopic Orbits of Subsystems in  Multiple
  Stars. IX}

\author{Andrei Tokovinin}
\affiliation{Cerro Tololo Inter-American Observatory | NSF's NOIRLab
Casilla 603, La Serena, Chile}
\email{andrei.tokovinin@noirlab.edu}

\begin{abstract}
New spectroscopic  orbits of  inner subsystems  in 14  hierarchies are
determined  from   long-term  monitoring   with  the   optical  echelle
spectrometer,  CHIRON.  Their  main components  are nearby  solar-type
stars belonging to nine triple systems (HIP 3645, 14307, 36165, 79980,
103735, 103814,  104440, 105879,  109443) and  five quadruples  of 2+2
hierarchy  (HIP 41171,  49336, 75663,  78163, and  117666). The  inner
periods range from  254 days to 18 yr.  Inner  subsystems in HIP 3645,
14313,  79979, 103735,  104440,  and 105879  are  resolved by  speckle
interferometry, and their  combined spectro-interferometric orbits are
derived here.   Astrometric orbits of  HIP 49336 Aa,Ab and  HIP 117666
Aa,Ab are determined  from wobble in the observed motion  of the outer
pairs.  Comparison with  three spectroscopic orbits found  in the Gaia
DR3 archive  reveals that Gaia under-estimated  the amplitudes (except
for  HIP 109443),  while the  periods match  approximately. This  work
contributes  new  data  on  the architecture  of  nearby  hierarchical
systems, complementing their statistics.

\end{abstract}



\section{Introduction}
\label{sec:intro}

Observations of  spectroscopic subsystems  in nearby  solar-type stars
are  motivated by  the  desire  to determine  their  periods and  mass
ratios,   complementing  statistics   of  hierachies   in  the   solar
neighborhood  \citep{FG67b}.   Many  subsystems discovered  by,  e.g.,
\citet{N04} or  by astrometric acceleration lack  orbits and therefore
confuse the statistics.  A long-term program at the 1.5 m telescope at
Cerro  Tololo   with  the   CHIRON  high-resolution   optical  echelle
spectrograph has been conducted to determine the missing periods, with
the  goal to  reach  relative completeness  for  periods shorter  than
$\sim$1000 days.  The  results obtained so far were  reported in eight
papers; the  last paper 8  \citep{chiron8} contains references  to the
full  series.  The  total  number of  spectroscopic orbits  determined
throughout this program  is 102.  Summary and  statistical analysis of
this  material  are presented  in  the  accompanying paper  10  (2022,
submitted).   The first  papers resulting  from this  project featured
short-period orbits, but longer periods  became accessible as the time
coverage increased.   Most orbits  presented here have  periods longer
than a  year.  Some  of them are  preliminary, lacking  adequate phase
coverage,  but  they  are   still  useful  for  statistical  purposes,
justfying their publication here. Six inner subsystems are wide enough
to be resolved by speckle  interferometry, allowing calculation of the
combined spectro-interferometric orbits.

In 2022 June, the third release of the Gaia catalog (GDR3) has changed
the  landscape by  publishing  $\sim10^5$ spectroscopic  orbits and  a
comparable  number  of astrometric  orbits  in  their non-single  star
catalog,  NSS \citep{Arenou2022}.   However, stars  with close  visual
companions were  removed from the  Gaia SB sample.  Comparison  of the
NSS with the CHIRON orbits, presented in paper 10, shows an overlap of
only  about 30\%,  so the  NSS completeness  with respect  to multiple
stars  remains  low.  Some  NSS  orbits  in  common with  CHIRON  have
substantially  different   parameters  (examples  are   found  below).
Although the NSS orbits contribute  significantly to the statistics of
nearby  hierarchies,   they  do  not  yet   replace  the  ground-based
monitoring and do not render the CHIRON survey obsolete.

This paper is organized similarly to  the previous ones.  The data and
methods  are  outlined  in Section~\ref{sec:orb},  where  the  orbital
elements are  also given.  The  hierarchical systems are  discussed in
Section~\ref{sec:obj}.   A   short  summary   in  Section~\ref{sec:sum}
concludes the paper.

\section{New Spectroscopic Orbits}
\label{sec:orb}

\begin{deluxetable*}{c c rr   l cc rr r c }
\tabletypesize{\scriptsize}     
\tablecaption{Basic Parameters of Observed Multiple Systems
\label{tab:objects} }  
\tablewidth{0pt}                                   
\tablehead{                                                                     
\colhead{WDS} & 
\colhead{Comp.} &
\colhead{HIP} & 
\colhead{HD} & 
\colhead{Spectral} & 
\colhead{$V$} & 
\colhead{$V-K_s$} & 
\colhead{$\mu^*_\alpha$} & 
\colhead{$\mu_\delta$} & 
\colhead{RV} & 
\colhead{$\varpi$\tablenotemark{a}} \\
\colhead{(J2000)} & 
 & &   &  
\colhead{Type} & 
\colhead{(mag)} &
\colhead{(mag)} &
\multicolumn{2}{c}{ (mas yr$^{-1}$)} &
\colhead{(km s$^{-1}$)} &
\colhead{(mas)} 
}
\startdata
00467$-$0426    & A  & 3645   & 4449  & G5      & 7.58 &  2.00 & 24*   & $-$261* & 9.7   & 30.08: \\
                & B  & \ldots & \ldots & M4V    & 15.20 & 4.92 & 21    & $-$260 & \ldots & 30.51 \\  
03046$-$5119    & A  & 14307  & 19330  & F8V    & 7.54  & 1.25 & 88    & 71     & 20.4   & 18.36 \\
                & B  & 14313  & \ldots & K1V    & 8.59  & 1.91 & 85    & 72     & 20.2   & 18.42: \\
07270$-$3419    & A  & 36165  & 59099  & F6V    & 7.03  & 1.23 & $-$305* & 96*  & 65.4   & 20.32: \\
                & B  & 36160  & 59100  & G1.5V  & 8.19  & 1.59 &  $-$307 & 91   & 64.9   & 20.71 \\ 
08240$-$1548    & AB & 41171  & 70904  & F2/F3V & 8.55  & 1.06 & $-$28* & $-$16*& $-$1.4 & 4.94: \\ 
10043$-$2823    & A  & 49336  & 87416  & F6V    & 7.82  & 1.19 & $-$27 & $-$23  & $-$13.4 & 10.67: \\
                & B  & \ldots & \ldots & \ldots & 8.19 & \ldots & $-$49 & $-$36  & $-$11.8 & 10.94: \\
15275$-$1058    & A  & 75663  & 137613 & G0     & 8.14 & 1.35   &  $-$62* & $-$36* &  $-$56.3 & 7.73 \\
                & B  &  \ldots & \ldots & \ldots & 9.21 & 1.50  & $-$61 & $-$35  &  $-$56.8 & 7.78 \\
15577$-$3915    & A  & 78163   & 142728 & G3/5V  & 9.04 & 1.54  &   17   & 7     & 9.4      & 10.49 \\
                & B  &  \ldots & \ldots & \ldots & 10.30 & 2.08 &   18   & 7     & 10.5     & 10.65 \\ 
16195$-$3054    & A  & 79980  & 146836  & F5IV  &  5.51  & 1.14 &   82   & 23    & 0.3      & 22.71 \\
                & B  & 79979  & 146835  & F9V   & 6.82   & 1.11 &   76   & 27    & $-$0.9   & 25.53: \\
21012$-$3511    & A  & 103735 & 1999918 & G3V   & 7.66   & 1.61 & $-$176 & $-$63 & 61.6     & 22.10: \\
                & B  & \ldots & \ldots & \ldots & 17.14  & 1.64 & $-$176 & $-$67 & \ldots   & 22.09 \\
21022$-$4300    & A  & 103814 & 200011 & G3IV+K0IV   & 6.64   & 1.62 &  71* & $-$112* & $-$33.5  & 11.25 \\
                & B  & 103819 & 200026 & K0III  & 6.90   & 2.27 &  70  & $-$111  & $-$35.6  & 11.25 \\
21094$-$7310    & AB & 104440 & 200525 & F9.5V  & 5.68  &  1.49 & 445*  & $-$330* & $-$11.1 & 46.99: \\
                & C  & \ldots & \ldots & \ldots & 13.50 &  6.16 & 433  & $-$303  & $-$8.3   & 50.6  \\
21266$-$4604    & A  & 105879 & 203934 & F7V    & 7.18  &  1.28 &  29*  & $-$112* & 35.5    & 12.44: \\
                & D  & \ldots & \ldots & \ldots & 9.96  &  1.76 &  31  & $-$112  & 35.7     & 13.10 \\
22104$-$5158    & A  & 109443 & 210236 & F8V    & 7.63  &  1.32 & 220*  & $-$104* & $-$3.8   & 15.33: \\
                & B  &  \ldots & \ldots & \ldots& 13.25 & \ldots& 225   &  $-$104 & \ldots  & 15.58 \\
23518$-$0637    & AB & 117666 & 223688 & G5V    & 8.73  &  1.69 &  85* & $-$12*  &   14.3     & 13.4: \\
\enddata
\tablenotetext{}{Proper motions and parallaxes are 
  from Gaia DR3 \citep{gaia3}. Colons mark parallaxes biased by subsystems, asterisks mark  PMs 
 from \citet{Brandt2021}.  }
\end{deluxetable*}

The hierarchical     systems     studied     here    are     listed     in
Table~\ref{tab:objects}. The  data are collected from  Simbad and GDR3
\citep{gaia3},   the  radial   velocities   (RVs)  are   mostly
determined in this work. The  first column gives the Washington Double
Star \citep[WDS,][]{WDS} code based on the J2000 coordinates.  The HIP
and HD  identifiers, spectral types, photometric  and astrometric data
refer either to the individual stars or to the unresolved subsystems. 
Parallaxes potentially biased by unresolved subsystems are marked by
colons, and asterisks indicate proper motions from \citet{Brandt2021}.

\subsection{Spectroscopic Observations}

Observations, data reduction, and orbit calculations were described in
previous papers of this series \citep[e.g.][]{chiron8}. To avoid
repetition, only a brief outline is given here. 

The spectra used here were taken with the 1.5 m telescope sited at the
Cerro Tololo  Inter-American Observatory (CTIO) in  Chile and operated
by  the   Small  and   Medium  Aperture  Telescopes   Research  System
\href{http://www.astro.yale.edu/smarts/}{(SMARTS)}
Consortium.\footnote{ \url{http://www.astro.yale.edu/smarts/}} Fifteen
hours of observing  time were allocated to this  program per semester,
starting from 2017B.  Observations were made with the fiber-fed CHIRON
optical   echelle  spectrograph   \citep{CHIRON,Paredes2021}  by   the
telescope operators in service mode.  The spectra taken with the image
slicer have a resolution of 85\,000.  They are reduced by the standard
CHIRON  pipeline.    The  wavelength  calibration  is   based  on  the
thorium-argon lamp spectra taken after each object.

The RVs  are determined  from Gaussian  fits to  the cross-correlation
function (CCF) of echelle orders with the binary mask constructed from
the solar spectrum,  as detailed in \citet{chiron1}.    The  RV errors
depend on  several factors such as  the width and contrast  of the CCF
dip,  blending with  other  dips, and  signal-to-noise  ratio.  The  rms
residuals from  the orbits can be  as low as 0.02  \kms, but typically
are between 0.1  and 0.5 \kms for the systems  studied here.  I assign
the RV  errors (hence  weights) to match  roughly the  residuals, with
larger errors for blended or noisy dips.  Some blended CCFs are fitted
by fixing the  width or amplitude of  individual components determined
from other  spectra with  better-separated dips. Otherwise,  a heavily
blended dip is  fitted by a single Gaussian, and  the resulting biased
RV is assigned a large error and a low weight in the orbit fit.

The width of the CCF dip is related to the projected rotation velocity
$V \sin i$, while its area  depends on the spectral type, metallicity,
and, for  blended spectra  of several stars,  on the  relative fluxes.
Table~\ref{tab:dip} lists  average parameters  of the  Gaussian curves
fitted to the CCF dips.  It  gives the number of averaged measurements
$N$  (blended CCFs  of  double-lined binaries  are  ignored), the  dip
amplitude  $a$,  its  dispersion  $\sigma$,  the  product  $a  \sigma$
proportional to  the dip area  (hence to  the relative flux),  and the
projected rotation velocity $V \sin i$, estimated from $\sigma$ by the
approximate formula given  in \citep{chiron1} and valid  for $\sigma <
12$ \kms.   The last column indicates  the presence or absence  of the
lithium 6708\,\AA ~line in individual components.

\begin{deluxetable*}{l l c cccc c}    
\tabletypesize{\scriptsize}     
\tablecaption{CCF Parameters
\label{tab:dip}          }
\tablewidth{0pt}                                   
\tablehead{                                                                     
\colhead{HIP} & 
\colhead{Comp.} & 
\colhead{$N$} & 
\colhead{$a$} & 
\colhead{$\sigma$} & 
\colhead{$a \sigma$} & 
\colhead{$V \sin i$ } & 
\colhead{Li}
\\
 &  &  & &
\colhead{(km~s$^{-1}$)} &
\colhead{(km~s$^{-1}$)} &
\colhead{(km~s$^{-1}$)} &
\colhead{  6708\AA}
}
\startdata
3645      & Aa &  10 & 0.403 & 3.66 & 1.47 & 2.4 & N \\
3645      & Ab &  10 & 0.115 & 3.93 & 0.45 & 3.5 & N \\
14313     & Ba & 8   & 0.297 & 3.99 & 1.19 & 3.8 & N \\
14313     & Bb & 8   & 0.232 & 4.30 & 1.00 & 4.7 & N \\
36165     & Aa & 9   & 0.208 & 5.19 & 1.08 & 7.0 & N \\ 
36160     & B  & 4   & 0.403 & 3.49 & 1.41 & 1.4 & N \\ 
41171     & Aa & 12  & 0.030 & 18.08 & 0.46 & 32.0 & N \\ 
41171     & Ab & 12  & 0.062 & 5.63 & 0.35  & 8.1 & N \\
41171     & Ba & 12  & 0.030 & 4.35 & 0.13  & 4.9 & N \\
41171     & Bb & 12  & 0.017 & 3.80 & 0.07  & 3.1 & N \\
49336     & Aa & 19  & 0.116 & 4.06 & 0.47  & 4.0 & N \\
49336     & Ab & 19  & 0.082 & 5.96 & 0.49  & 8.8 & N \\
49336     & B  & 19  & 0.028 & 6.97 & 0.19  & 10.9 & N \\
75663    & Aa & 19  & 0.218  & 6.85 & 1.50 & 10.7 & Y \\ 
78163    & Ba & 12  & 0.355  & 4.26 & 1.51 & 4.6  & N?  \\ 
79979    & Ba & 5   & 0.194  & 4.30 & 0.83 & 4.7  & Y \\  
79979    & Bb & 5   & 0.115  & 3.71 & 0.43 & 2.7  & N \\  
103735   & Aa  & 4  & 0.320  & 3.59 & 1.15 & 2.1 &  Y \\
103735   & Ab  & 4  & 0.058  & 3.83 & 0.22 & 3.2 & N  \\
104440   & Aa  & 3  & 0.274  & 5.03 & 1.38 & 6.7 & Y \\
104440   & Ab  & 3  & 0.025  & 4.96 & 0.13 & 6.5 & N \\
103814   & Aa  & 8  & 0.297  & 3.92 & 1.16 & 3.5 & N  \\
105879   & Aa  & 1  & 0.193  & 5.69 & 1.10 & 6.9? & N \\
105879   & Ab  & 1  & 0.045  & 3.19 & 0.14 & 1.0? & N \\
109443   & Aa  & 6  & 0.266  & 4.97 & 1.32 & 6.5 & Y \\
117666   & Aa  & 10 & 0.209  & 3.81 & 0.80 & 3.1 & N? \\
117666   & Ba  & 10 & 0.194  & 3.69 & 0.71 & 2.6 & N? \\
\enddata 
\end{deluxetable*}

\begin{deluxetable*}{l l cccc ccc c c}    
\tabletypesize{\scriptsize}     
\tablecaption{Spectroscopic Orbits
\label{tab:sborb}          }
\tablewidth{0pt}                                   
\tablehead{                                                                     
\colhead{HIP} & 
\colhead{System} & 
\colhead{$P$} & 
\colhead{$T$} & 
\colhead{$e$} & 
\colhead{$\omega_{\rm A}$ } & 
\colhead{$K_1$} & 
\colhead{$K_2$} & 
\colhead{$\gamma$} & 
\colhead{rms$_{1,2}$} &
\colhead{$M_{1,2} \sin^3 i$} 
\\
& & \colhead{(d)} &
\colhead{(JD -2,400,000)} & &
\colhead{(deg)} & 
\colhead{(km~s$^{-1}$)} &
\colhead{(km~s$^{-1}$)} &
\colhead{(km~s$^{-1}$)} &
\colhead{(km~s$^{-1}$)} &
\colhead{ (${\cal M}_\odot$) } 
}
\startdata
3645 & Aa,Ab & 1529.6 & 59055.5 & 0.240       & 176.3    &  9.985      & 11.423    & 9.597      & 0.037 & 0.78  \\
     &    & $\pm$3.7 & $\pm$9.5  & $\pm$0.012 & $\pm$2.3 & $\pm$0.271 & $\pm$0.275 & $\pm$0.132 & 0.145 & 0.68 \\
14313 & Ba,Bb & 6648.1 & 53347.9    & 0.488      & 318.1   & 8.048       & 8.228     & 21.540      & 0.184 & 1.01 \\
      &   & $\pm$96.6 & $\pm$105.7 & $\pm$0.008 & $\pm$1.5 & $\pm$0.153 & $\pm$0.155 & $\pm$0.049  & 0.191  & 0.99 \\
36165 & Aa,Ab & 2300.4 & 57921.4   & 0.610      & 64.4     & 5.69      & \ldots & 65.52     & 0.037  & 1.28:\\
      &   & $\pm$16.1  & $\pm$37.9 & $\pm$0.075 & $\pm$4.9 & $\pm$1.16 & \ldots & $\pm$0.34 & \ldots & 0.39 \\
41171 & Ba,Bb &  963.1 & 58913.2 & 0.607        & 273.4    & 15.61     & 18.70     & $-$3.32    & 0.30 & 1.10  \\
      &   & $\pm$1.7 & $\pm$1.7 & $\pm$0.007   & $\pm$0.8  & $\pm$0.16 & $\pm$0.25 & $\pm$0.07  & 0.46 & 0.92 \\
41171 & Aa,Ab &  25.4133 & 58449.999 & 0.5320     & 308.25    & 47.18     & 48.27     & $-$1.41    & 1.84  & 0.70  \\
      &   & $\pm$0.0001 & $\pm$0.005 & $\pm$0.0005 & $\pm$0.09  & $\pm$0.23 & $\pm$0.03 & $\pm$0.03  & 0.13 & 0.69 \\
49336 & Ba,Bb & 1307.4 & 58895.8 & 0.163       & 132.5     & 4.55      & \ldots   & $-$11.73 &  0.19  & 1.35: \\
      &   & $\pm$8.4 & $\pm$38.4 & $\pm$0.021  & $\pm$11.4 & $\pm$0.15 & \ldots   & $\pm$0.09& \ldots  & 0.50 \\
75663 & Aa,Ab & 623.76 & 59098.35 & 0.653      & 269.9     & 10.045     & \ldots &  $-$56.352 &  0.041 & 1.47: \\
      &  & $\pm$0.21 & $\pm$0.47 & $\pm$0.002  & $\pm$0.6  & $\pm$0.053 & \ldots & $\pm$0.029 & \ldots & 0.48 \\
78163 & Ba,Bb  & 2083.2 & 59208.6 & 0.619      & 27.7      & 10.84     & \ldots  & 10.55       & 0.080 & 0.93: \\
      &   & $\pm$20.4 & $\pm$20.0 & $\pm$0.025 & $\pm$4.9  & $\pm$1.63 & \ldots  & $\pm$0.45   & \ldots & 0.71 \\
79979 & Ba,Bb & 1083.16 & 57635.32 & 0.610     & 349.8     & 15.31    & 18.01    &  0.00      & 0.121 & 1.14 \\
      &   & $\pm$1.84 & $\pm$4.81 & $\pm$0.003 & $\pm$1.3 & $\pm$0.16 & $\pm$0.21& $\pm$0.06  & 0.204 & 0.97 \\
103735 & Aa,Ab & 4251.8 & 59433.5 & 0.368      & 152.4    & 7.29      & 10.13    & 61.69      &  0.020 & 1.09 \\
       &  & $\pm$12.1 & $\pm$11.3 & $\pm$0.003 & $\pm$1.2 & $\pm$0.04 & $\pm$0.23 & $\pm$0.05 & 0.375  & 0.79 \\
103814 & Aa,Ab & 1089.8 & 58393.5 & 0.601      & 331.0    & 4.48      & \ldots & $-$33.62    & 0.010  & 1.78: \\
       &  & $\pm$9.4 & $\pm$16.1  & $\pm$0.079 & $\pm$9.1 & $\pm$1.44 & \ldots & $\pm$0.23   & \ldots & 0.28 \\
104440 & A,B & 1947.5 & 57909.2 & 0.631        & 178.5    & 10.202    & 16.759   & $-$11.211    & 0.020 & 1.15 \\
     &    & $\pm$0.9 & $\pm$1.7 & $\pm$0.002  & $\pm$0.8  & $\pm$0.044 & $\pm$0.180 & $\pm$0.046& 0.494 & 0.70 \\
105879 & Aa,Ab & 2935.6 & 60032.2  & 0.631     & 359.7     & 11.07   & 13.83       & 35.44    & 0.418 & 1.24 \\
       &  & $\pm$9.4   & $\pm$13.0 & $\pm$0.019 & $\pm$2.5 & $\pm$0.66 & $\pm$0.91 & $\pm$0.14 & 0.843 & 1.00 \\
109443 & Aa,Ab & 978.5 & 58715.3  & 0.214        & 24.3      & 3.02   & \ldots    & $-$3.80  & 0.008  & 1.30: \\
       &  & $\pm$37.4 & $\pm$210.9 &  $\pm$0.084 & $\pm$67.7 & $\pm$0.34 & \ldots & $\pm$0.27 & \ldots & 0.18 \\
117666 & Aa,Ab & 781.2 & 59353.7 & 0.105         & 128.4     & 4.944      & \ldots & 14.336 & 0.089   & 0.97: \\
       &  & $\pm$1.6 & $\pm$62.8 & $\pm$0.042    & $\pm$29.4 & $\pm$0.135 & \ldots & $\pm$0.128& \ldots & 0.95: \\
117666 & Ba,Bb & 253.9 & 59204.8 & 0.269         & 113.9     & 9.28      & \ldots & 14.264   & 0.128   &  0.42 \\
       &  & $\pm$0.145 & $\pm$2.5 & $\pm$0.018   & $\pm$3/3  & $\pm$0.16 & \ldots & $\pm$0.132 & \ldots & 0.31 
\enddata 
\end{deluxetable*}

\begin{deluxetable}{r l c rrr c }    
\tabletypesize{\scriptsize}     
\tablecaption{Radial Velocities and Residuals (fragment)
\label{tab:rv}          }
\tablewidth{0pt}                                   
\tablehead{                                                                     
\colhead{HIP} & 
\colhead{System} & 
\colhead{Date} & 
\colhead{RV} & 
\colhead{$\sigma$} & 
\colhead{(O$-$C) } &
\colhead{Comp.}  \\
\colhead{HD} & & 
\colhead{(JD -2,400,000)} &
\multicolumn{3}{c}{(km s$^{-1}$)} &
\colhead{Instr.}
}
\startdata
  3645 &Aa,Ab  &  54781.5350 &    8.94 &    2.00 &    0.22 &  a \\
  3645 &Aa,Ab  &  57985.7810 &   13.38 &    0.50 &    0.05 &  a \\
  3645 &Aa,Ab  &  57985.7810 &    4.95 &    0.70 &   -0.40 &  b \\
  3645 &Aa,Ab  &  58130.5330 &   16.01 &    0.05 &   -0.01 &  a \\
  3645 &Aa,Ab  &  58130.5330 &    2.33 &    0.25 &    0.08 &  b 
\enddata 
\tablenotetext{}{(This table is available in its entirety in
  machine-readable form). Instrument codes:  
B -- \citet{Butler2017};
E -- Fiber echelle \citep{Tok2015};
F -- \citet{Frasca2018}
G -- Gaia DR2;
L -- DuPont echelle \citep{LCO};
N -- \citet{Nidever2002}
}
\end{deluxetable}


\begin{deluxetable*}{l l cccc ccc}    
\tabletypesize{\scriptsize}     
\tablecaption{Visual and Astrometric Orbits
\label{tab:vborb}          }
\tablewidth{0pt}                                   
\tablehead{                                                                     
\colhead{HIP} & 
\colhead{System} & 
\colhead{$P$} & 
\colhead{$T$} & 
\colhead{$e$} & 
\colhead{$a$} & 
\colhead{$\Omega_{\rm A}$ } & 
\colhead{$\omega_{\rm A}$ } & 
\colhead{$i$ }  \\
& & \colhead{(yr)} &
\colhead{(yr)} & &
\colhead{(arcsec)} & 
\colhead{(deg)} & 
\colhead{(deg)} & 
\colhead{(deg)} 
}
\startdata
3645   & Aa,Ab  & 4.188   & 2020.563     & 0.240       & 0.0967        & 314.0 &   176.3  &  97.6 \\
    &         & $\pm$0.010  & $\pm$0.026 & $\pm$0.012  & $\pm$0.0014   & $\pm$0.7  & $\pm$1.7  &  $\pm$1.0 \\  
14313  & Ba,Bb & 18.20   & 2004.94       & 0.488       & 0.1591       & 292.6      &  318.1   & 84.2   \\ 
    &         & $\pm$0.68  & $\pm$0.29  & $\pm$0.008 & $\pm$0.0017  & $\pm$0.3   & $\pm$1.6  & $\pm$0.5  \\  
49336   & Ba,Bb & 3.580   & 2020.127     & 0.163       & 0.0071     & 9.3     & 132.5    & 140.4   \\ 
     &         & $\pm$0.023 & $\pm$0.105 & $\pm$0.021  & $\pm$0.0009   & $\pm$6.0  & $\pm$11.4  &  $\pm$13.5 \\  
49336   & A,B & 397.8   & 1971.24        & 0.729      & 0.886      & 321.5     & 253.5    & 142.3   \\ 
     &         & $\pm$17.9 & $\pm$0.32   & $\pm$0.010 & $\pm$0.021  & $\pm$1.6  & $\pm$1.9  &  $\pm$1.5 \\  
79979 & Ba,Bb  & 2.966      & 2016.674   & 0.610      & 0.0601      & 103.8    & 349.8    & 82.9   \\
    &         & $\pm$0.005  & $\pm$0.013 & $\pm$0.003 & $\pm$0.0007& $\pm$0.7 & $\pm$1.3  &  $\pm$1.0 \\
103735 & Aa,Ab & 11.64   &   2021.598    & 0.368      & 0.1363      & 168.1     & 152.4    & 87.4 \\
    &         & $\pm$0.03 & $\pm$0.031  & $\pm$0.003  & $\pm$0.0009& $\pm$0.4  & $\pm$1.2  &  $\pm$0.7 \\
104440  & A,B  & 5.332     & 2017.424     & 0.631     & 0.1905       & 194.3    & 178.5   & 93.0   \\
     &         & $\pm$0.003 & $\pm$0.005 & $\pm$0.002& $\pm$0.0013& $\pm$0.4  & $\pm$0.8  &  $\pm$1.0 \\  
105879  & Aa,Ab & 8.037     & 2023.237   & 0.631      & 0.0669       & 231.5  & 359.7      & 97.0 \\    
    &         & $\pm$0.026  & $\pm$0.036 & $\pm$0.019 & $\pm$0.0018   & $\pm$1.0  & $\pm$2.5  &  $\pm$1.6 \\  
117666  & Aa,Ab &  2.138      & 2021.33 & 0.105      & 0.0079       & 26.1      & 128.4      & 148.0  \\
     &         & $\pm$0.004  & $\pm$0.17 & $\pm$0.043  & $\pm$0.0007    & $\pm$5.3  & $\pm$29.4  & fixed \\  
117666  & A,B  &  30.073      & 2020.392  & 0.309      & 0.1809       & 24.3      & 356.5      & 147.5  \\
     &         & $\pm$0.071  & $\pm$0.044 & $\pm$0.004  & $\pm$0.0009    & $\pm$1.4  & $\pm$1.5  & fixed \\  
\enddata 
\end{deluxetable*}

\begin{deluxetable}{r l r r }    
\tabletypesize{\scriptsize}     
\tablecaption{Radial Velocities of Other Components
\label{tab:rvconst}          }
\tablewidth{0pt}                                   
\tablehead{                                                                     
\colhead{HIP} & 
\colhead{Comp.} & 
\colhead{Date} & 
\colhead{RV}   \\ 
 & & 
\colhead{(JD -2,400,000)} &
\colhead {(km s$^{-1}$)}  
}
\startdata
14257 & C & 57986.8700 & 28.496 \\
36160 & B & 56940.8450 & 64.859 \\
36160 & B & 57266.9055 & 64.884 \\ 
36160 & B & 58121.7557 & 64.889 \\
36160 & B & 58193.5556 & 64.881 \\   
36160 & B & 58546.5656 & 64.873 \\
36160 & B & 59168.7635 & 64.899 \\
105879 & D & 55477.5014  & 36.077 \\   
105879 & D & 56885.7159  & 35.714 \\
\enddata 
\end{deluxetable}

\begin{deluxetable*}{c c r rrr rr }

\tabletypesize{\scriptsize}
\tablewidth{0pt}
\tablecaption{Positional Measurements and Residuals \label{tab:obs}}
\tablehead{
\colhead{HIP} & 
\colhead{System} & 
\colhead{$T$} &
\colhead{$\theta$} & 
\colhead{$\rho$} &
\colhead{$\sigma$} & 
\colhead{O$-$C$_\theta$} & 
\colhead{O$-$C$_\rho$} \\
& & 
\colhead{(yr)} & 
\colhead{(\degr)} &
\colhead{($''$)} & 
\colhead{($''$)} & 
\colhead{(\degr)} &
\colhead{($''$)} 
}
\startdata
  3645 &Aa,Ab & 2011.6850 & 159.8 &  0.0300 &   0.002 &   5.4 &   0.000 \\
  3645 &Aa,Ab & 2011.9417 & 139.9 &  0.0598 &   0.002 &  $-$0.1 &  $-$0.000 \\
  3645 &Aa,Ab & 2014.7537 & 311.2 &  0.1076 &   0.002 &   0.3 &  $-$0.000 \\
  3645 &Aa,Ab & 2021.8909 & 320.5 &  0.0742 &   0.002 &  $-$1.9 &  $-$0.001 \\
  3645 &Aa,Ab & 2022.4447 & 316.3 &  0.1153 &   0.002 &   0.2 &   0.000 \\
 14313 &Ba,Bb & 2014.7635 & 107.9 &  0.1911 &   0.002 &  $-$0.4 &   0.001 \\
 14313 &Ba,Bb & 2014.7635 & 108.3 &  0.1908 &   0.002 &  $-$0.0 &   0.000 \\
 \enddata
\tablenotetext{}{(This table is available in its entirety in
  machine-readable form) }
\end{deluxetable*}

\subsection{Orbit Calculation}

The  orbital  elements   and  their  errors  are   determined  by  the
least-squares fits with weights  inversely proportional to the adopted
RV    errors.    The    IDL   code    {\tt   ORBIT}\footnote{Codebase:
  \url{http://www.ctio.noirlab.edu/\~atokovin/orbit/}              and
  \url{https://doi.org/10.5281/zenodo.61119}      }      was      used
\citep{orbit}. Several  double-lined pairs studied here  were resolved
by speckle  interferometry, and in  such case the combined  orbits are
fitted jointly to  the RVs and position measurements.   In some triple
systems,  the orbits  of the  outer  and inner  subsystems are  fitted
jointly to the RVs and, where available, position measurements using a
modification of the same code {\tt ORBIT3} \citep{ORBIT3} described by
\citet{TL2017}. Both codes allow to fix some orbital elements to avoid
degeneracies (e.g.  for circular  or face-on orbits)  or to  cope with
insufficient data (e.g.  an incomplete coverage of the outer orbit).

Table~\ref{tab:sborb} gives  elements of  the spectroscopic  orbits in
standard notation.  Its  last column contains the masses  $M \sin^3 i$
for double-lined binaries.  For single-lined  systems, the mass of the
primary star (listed  with colons) is estimated from  its absolute $V$
magnitude, and the  minimum mass of the secondary  that corresponds to
the    90\degr   ~inclination    is    derived    from   the    orbit.
Table~\ref{tab:rv},   published  in   full  electronically,   provides
individual RVs and  residuals to orbits.  The Hipparcos  number of the
primary star and the system identifier (components joined by comma) in
the first two  columns define the pair.  Then follow  the Julian date,
the  RV, its  adopted error  $\sigma$ (blended  CCF dips  are assigned
larger  errors), and  the residual  to  the orbit  (O$-$C).  The  last
column  specifies to  which  component  this RV  refers  ('a' for  the
primary,  'b'  for the  secondary).   The  RVs  of some  other  visual
components are provided, for completeness, in Table~\ref{tab:rvconst}.
It contains the HIP number, the component letter, the Julian date, and
the RV. The less accurate RVs  derived from blended dips are marked by
colons.

The elements of visual orbits  are given in Table~\ref{tab:vborb}. For
combined spectro-interferometric  orbits, it repeats  common elements,
but the period $P$ and epoch $T$ are given in Julian years rather than
days. This  table also  contains elements of  the outer  visual orbits
fitted  jointly with  the inner  subsystems using  {\tt ORBIT3}.   The
positional measurements used in these orbits are published (except the
latest  observations  at SOAR);  they  are  listed together  with  the
adopted errors and residuals  in Table~\ref{tab:obs}.

\subsection{Complementary Data}

I use here  astrometry and photometry from the  GDR3 \citep{gaia3} and
from the  earlier data releases  where needed.  For  multiple systems,
the  standard   astrometry  is  compromised  by   acceleration  and/or
unresolved  companions  (this  bias  is reduced  for  the  stars  with
astrometric solutions  in the NSS).  The RUWE parameter  (Reduced Unit
Weight  Error) captures  the excessive  astrometric noise,  helping to
identify biased  astrometry in  GDR3.  Most (but  not all)  stars with
subsystems studied here have  RUWE$>2$.  Uncertain Gaia parallaxes are
marked by  colons in Table~\ref{tab:objects}.   Astrometric subsystems
are detected by  the increased RUWE or by the  difference $\Delta \mu$
between the  short-term proper  motion (PM) measured  by Gaia  and the
long-term  PM $\mu_{\rm  mean}$ deduced  from the  Gaia and  Hipparcos
positions \citet{Brandt2021}.  For stars with  a large RUWE, I use the
long-term PMs  determined by Brandt  in place  of the PMs  measured by
Gaia.

For   some   systems,   spectroscopy  is   complemented   by   speckle
interferometry of the outer pairs. Most speckle observations used here
were made at the Southern  Astrophysical Research Telescope (SOAR) and
are  referred to  in  the  text simply  as  'SOAR  data'.  The  latest
observations  and  references  to  older  publications  are  found  in
\citet{Tokovinin2022}. Apart  from the position  measurements, speckle
interferometry provides differential photometry of close visual pairs.

\section{Individual Objects}
\label{sec:obj}

Figures in  this section show  the RV  curves and the  matching visual
orbits for resolved subsystems.  In the RV plots, green squares denote
the primary component, blue  triangles denote the secondary component,
while the full and dashed lines plot the orbit. Typical error bars are
smaller than  the symbols. In  the visual orbit plots,  squares denote
the  measured positions,  connected by  short lines  to the  ephemeris
positions on  the orbital ellipse  (solid line).   Masses of  stars are
estimated  from  absolute   magnitudes  using  standard  main-sequence
relations from \citet{Pecaut2013}.  Orbital  periods of wide pairs are
evaluated    statistically    from   their    projected    separations
\citep[see][]{MSC}.   Semimajor axes  of spectroscopic  subsystems are
computed using the third Kepler's  law, and the photocenter amplitudes
are evaluated based on the estimated masses and fluxes.

\subsection{HIP 3645 (Triple)}

\begin{figure}[ht]
\plotone{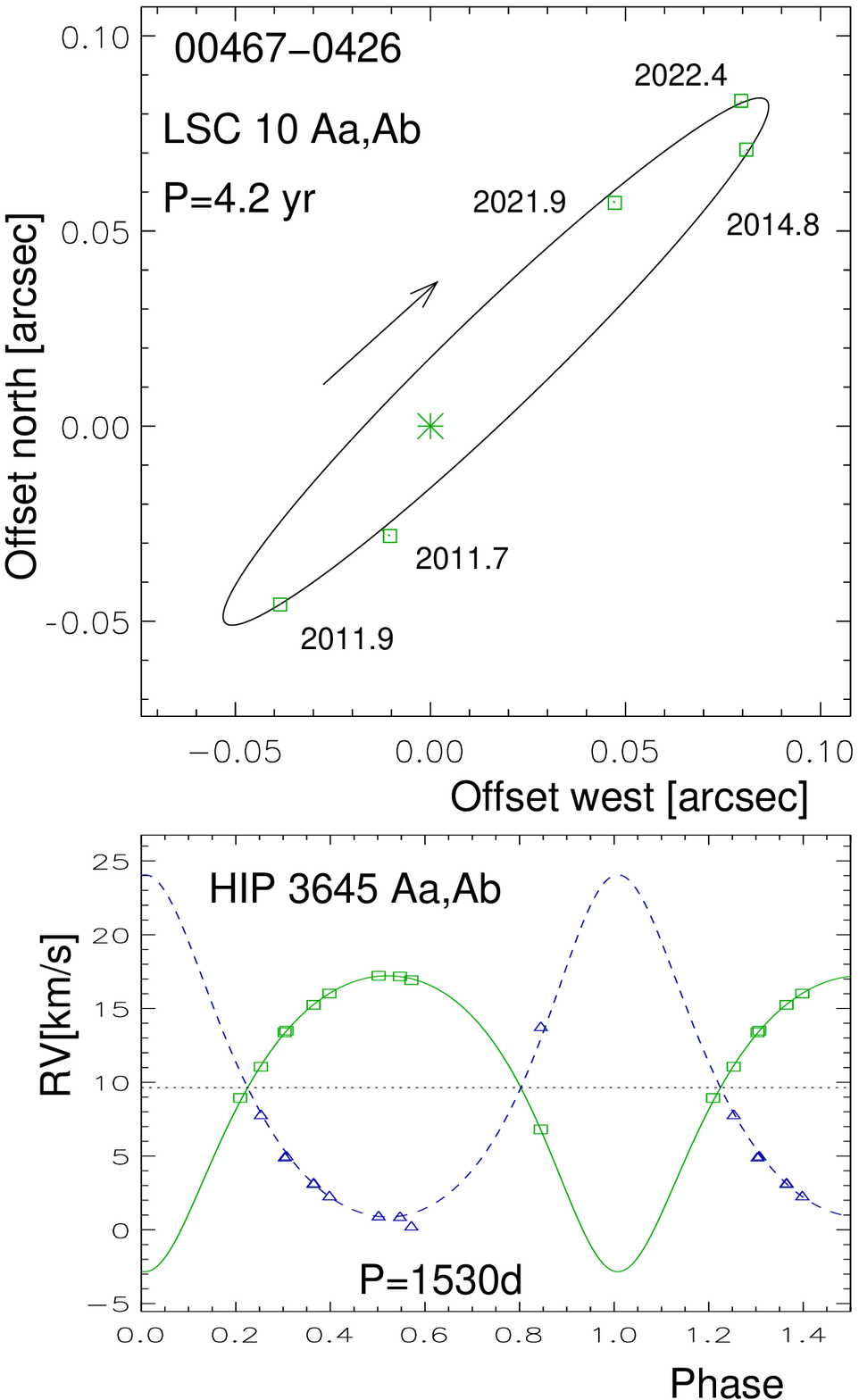}
\caption{Visual orbit and RV curve of HIP 3645 Aa,Ab.
\label{fig:3645} 
}
\end{figure}

This solar-type triple system belongs  to the 67-pc sample.  The outer
60\arcsec  ~common proper  motion (CPM)  pair A,B  (LDS~9100) has  been
discovered by \citet{Luyten1979}.  Star B  is an M4V dwarf of $V=15.2$
mag known  as LP~646-9 with  an accurate GDR3 parallax  of 30.51\,mas.
The parallax of  A is biased by the inner  subsystem, first discovered
as a  3.5 yr astrometric binary in  Hipparcos \citep{Goldin2007}. This
pair has been resolved in 2011 by \citet{Horch2017} at a separation of
30  mas (LSC~10  Aa,Ab).  Its  speckle monitoring  at SOAR  started in
2015.  The three first observations did  not resolve the pair, but the
measurements  in 2021  and  2022  are good,  indicating a  magnitude
difference of $\Delta I = 1.1$ mag and a separation of up to 0\farcs11.

Double  lines  were noted  by  \citet{N04},  and  the star  is  called
``Spectroscopic binary'' in Simbad.  Most CHIRON spectra of A are also
double-lined. The  RVs of  Aa and  Ab are used  here jointly  with the
position     measurements    to     derive     a    combined     orbit
(Figure~\ref{fig:3645}).   The  period  is  4.2 yr,  longer  than  the
Goldin's one.  The orbit  is oriented edge-on,  and the  RV amplitudes
translate  into Aa  and Ab  masses of  0.78 and  0.68 \msun,  somewhat
smaller  than  0.94  and  0.80   \msun  estimated  from  the  absolute
magnitudes.   The  visual  orbit,  unbiased parallax  of  B,  and  the
spectroscopic mass  ratio correspond  to the masses  of 0.97  and 0.85
\msun that  agree better with  the photometric estimates.   The masses
imply  RV  amplitudes  7\%  larger   than  measured,  and  this  minor
discrepancy could be  caused by line blending. The ratio  of dip areas
corresponds to $\Delta m_{\rm Aa,Ab} = 1.29$ mag, slightly larger than
measured by speckle in the $I$ band. Both components rotate slowly.

\subsection{HIP 14307+14313 (Triple)}

\begin{figure}[ht]
\plotone{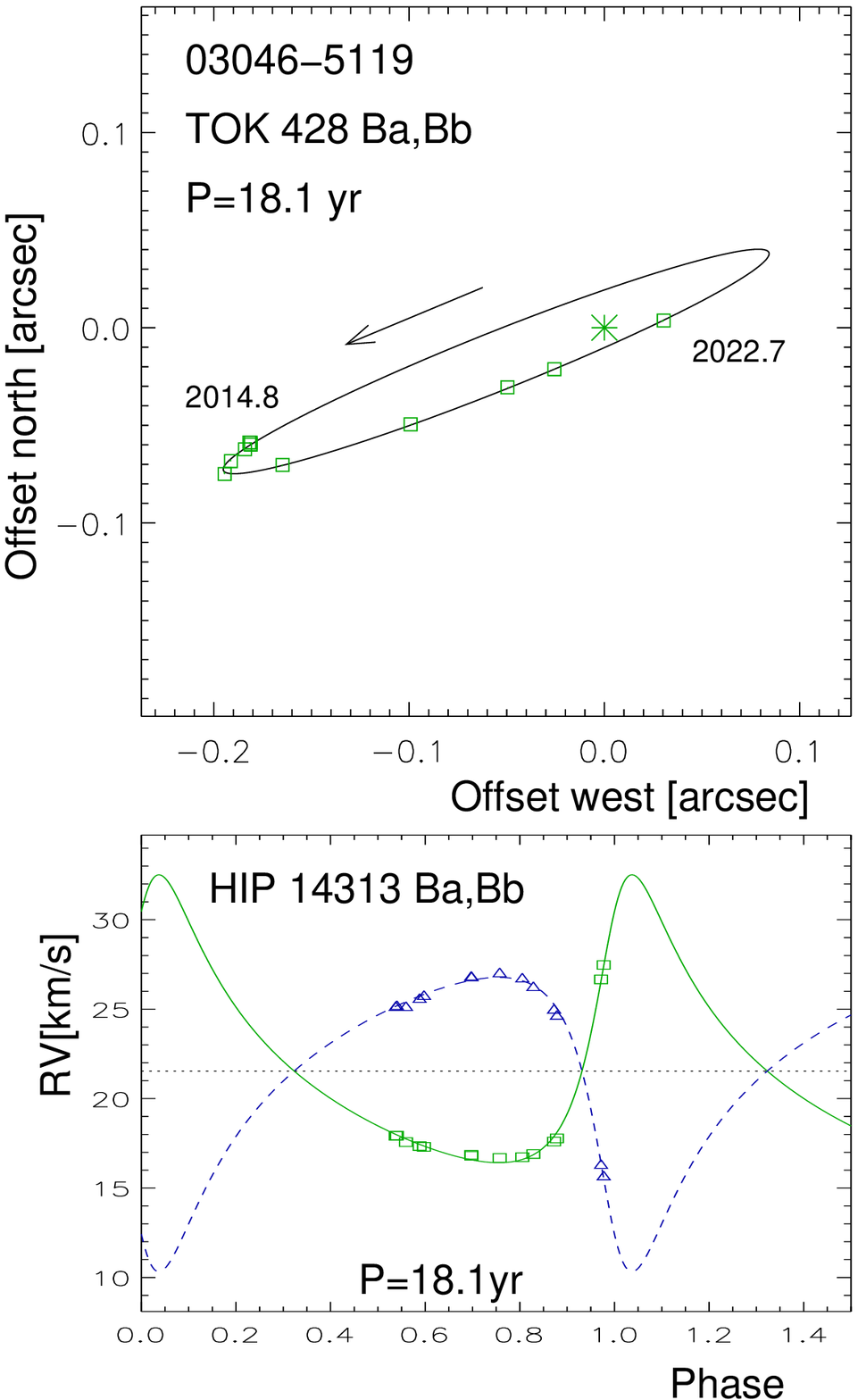}
\caption{Visual orbit and RV curve of HIP 14313 Ba,Bb.
\label{fig:14313} 
}
\end{figure}

This system has  some similarity to the previous one:  a wide binary
within 67\,pc  that hosts a  subsystem.  The 38\arcsec ~pair  A,B (DUN
10) has  been known  since 1826.   The WDS lists  another pair  A,C at
510\arcsec ~separation  (TOK~428). Star  C (HIP~14257,  HD~19254, F7V)
has a PM  of (98.8, 67.6) \masyr, similar  to the PMs of A  and B, and
for  this   reason  it  was  listed   in  the  survey  of   CPM  pairs
\citep{TokLep2012}.   However,  the  difference of  the  PM,  parallax
(14.70\,mas   according   to   GDR3),   and   RV   (28.5   \kms,   see
Table~\ref{tab:rvconst}) of  star C with respect  to A and B  rule out
its physical association.  There is  no excessive astrometric noise in
A and C (RUWE close to 1) in GDR3, and no parallax for B because it is
a close binary.

Star B has been resolved by speckle interferometry at SOAR in 2014 and
bears the  name TOK~428 Ba,Bb in  the WDS.  The pair  slowly opened up
from 0\farcs19  to 0\farcs21  by 2016, closed  down to  33\,mas in
2021.96,  and was  resolved  again in  2022.68  after passing  through
the conjunction. Double lines in the CHIRON  spectra  show only a slow
evolution, as  can be seen  in the RV  curve (Figure~\ref{fig:14313}).
The preliminary  combined 18 yr orbit  fitted to the RVs  and position
measurements predicts  periastron in 2023  March, when the  largest RV
difference between Ba  and Bb will occur.   Continued observations are
needed to improve our first orbit.

Speckle interferometry at SOAR established the magnitude difference of
$\Delta I_{\rm  Ba,Bb} = 0.42  \pm 0.06$ mag,  while the ratio  of dip
areas gives $\Delta m_{\rm Ba,Bb} = 0.19$ mag. The latter leads to the
visual magnitudes  of 9.25 and 9.44  for Ba and Bb,  respectively, and
the   ``photometric''   masses   of   0.90   and   0.87   \msun.   The
``spectroscopic'' masses are slightly larger, 1.01  and 0.99 \msun,
and the orbital parallax of  18.26\,mas matches well the GDR3 parallax
of A, 18.36\,mas. Stars Ba and Bb rotate slowly and have no lithium
line.

\subsection{HIP 36165+36160  (Triple)}

\begin{figure}[ht]
\plotone{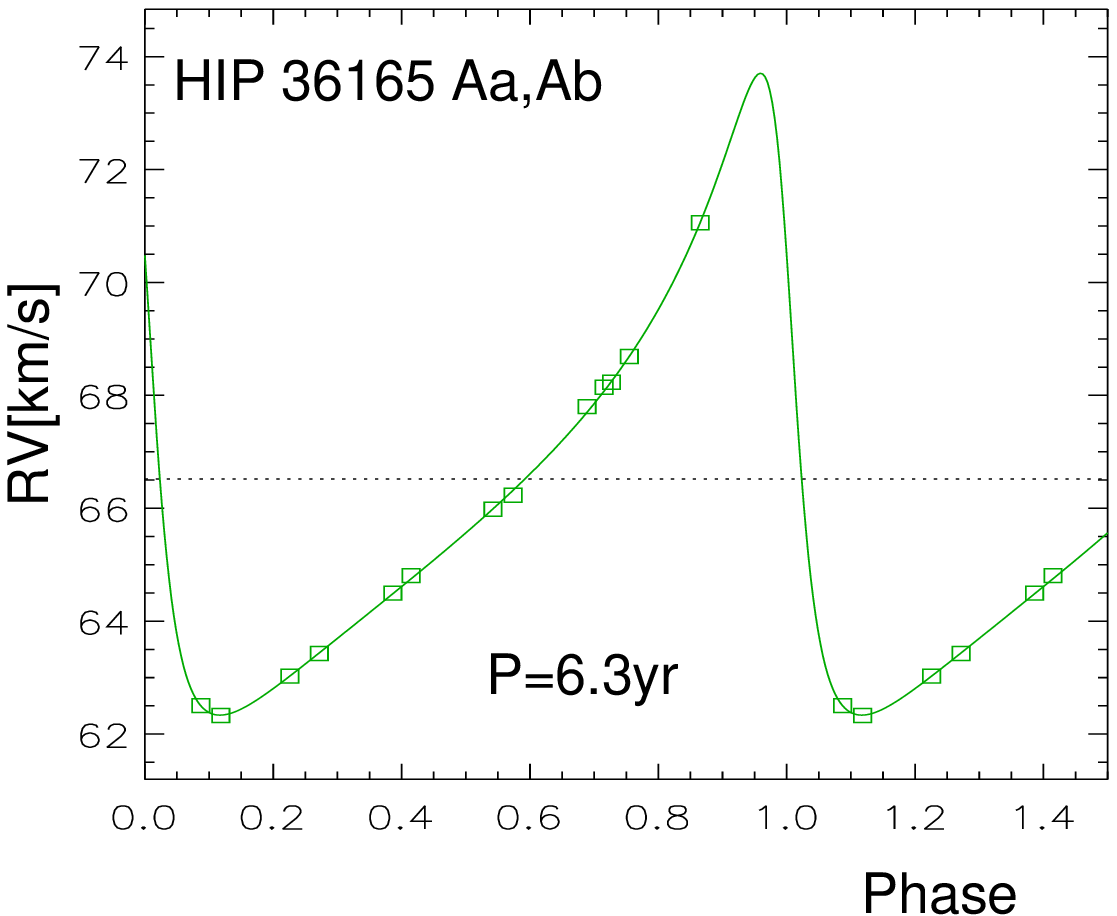}
\caption{The RV curve of HIP 36165 Aa,Ab. 
\label{fig:36165} 
}
\end{figure}

The wide 17\arcsec ~pair of HIP~36165  (star A, $V=7.03$ mag, F6V) and
HIP~36160 (star  B, $V=8.19$ mag,  G1.5V) has been discovered  by John
Herschel  in  1835  (HJ~3969).   The  fast  and  common  PM,  matching
parallaxes  and RVs  prove  the  bound nature  of  this  pair with  an
estimated period of 15 kyr.  \citet{N04}  noted that the RVs of both A
and B were variable.  However, CHIRON and other sources indicate that B
has a constant RV of 64.9 \kms  and is most likely a single star (RUWE
1.0 in  GDR3). The RV  of A,  on the other  hand, varies with  a small
amplitude;  this motion produces  astrometric noise in  Gaia (RUWE
11.9) and a large acceleration detected by \citet{Brandt2021}. 

The  spectroscopic orbit  with a  period of  6.3 yr  derived from  the
CHIRON RVs  is illustrated in Figure~\ref{fig:36165}.   The descending
part  of  the   RV  curve  is  not  yet  covered,   so  the  orbit  is
preliminary. The period is well  constrained, but the eccentricity can
be larger.  The estimated mass of Aa, 1.28 \msun, matches its spectral
type F6V. The minimum  mass of Ab is 0.39 \msun;  no spectral lines of
Ab are  detectable, while speckle  and adaptive optics imaging  \citep{Tok2010} has
not resolved any subsystems around stars A and B.

\subsection{HIP 41171  (Quadruple)}

\begin{figure}[ht]
\plotone{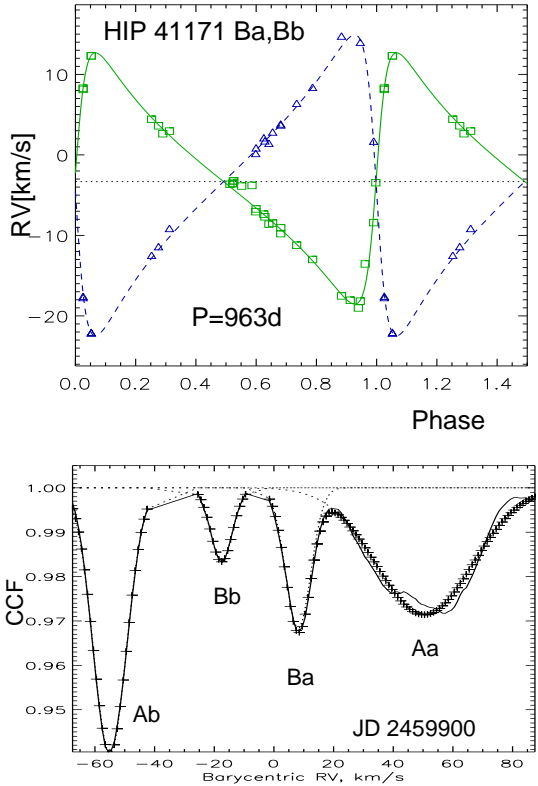}
\caption{The RV curve  of HIP 41171 Ba,Bb (top) and  the CCF with four
  well-separated dips recorded on JD 2,459,900 (bottom, solid line). The
  plus signs show the sum of four Gaussian curves, plotted individually
  by dotted lines.
\label{fig:41171} 
}
\end{figure}

This is a rare case of  a quadruple-lined object (SB4). The system has
been   presented   and  discussed   in   paper   6  of   this   series
\citep{chiron6}, where a 25-day SB2  orbit of Aa,Ab (main component in
the 0\farcs9  visual pair RST~4396)  was determined.  The lines  of Ba
and Bb  are clearly  separated from  the lines  of Aa  and Ab  only in
certain phases  of the 25-day  orbit.  Systematic monitoring  at these
moments during several  years has led eventually  to the determination
of the 2.6  yr orbit of Ba,Bb (Figure~\ref{fig:41171}). The  RVs of Aa
and Ab match  the published orbit; slightly refined  elements of Aa,Ab
derived with additional data are given in Table~\ref{tab:sborb}.

The visual magnitudes of Ba and Bb (10.95 and 11.38 mag, respectively)
were estimated from the areas of the four CCF dips. They correspond to
the masses of  1.09 and 1.00 \msun (mass  ratio $q_{\rm Ba,Bb}=0.92$),
similar to  the spectroscopic  masses $M  \sin^3 i$  of 1.10  and 0.92
\msun ($q_{\rm  Ba,Bb} = 0.83$).  This  means that the orbit  of Ba,Bb
has  a  large inclination.   It  is  oriented unfavorably  ($\omega  =
273\degr$), so  despite the  estimated semimajor  axis of  12\,mas the
pair Ba,Bb has never been  resolved by speckle interferometry at SOAR,
not even partially, in 9 visits.   The prospect of its resolution with
larger telescopes or interferometers is  good, though.  The outer pair
A,B  has an  estimated period  of  1.2 kyr  and moves  very slowly  in
retrograde sense. It has covered a 22\degr ~arc since its discovery in
1940.  The 0\farcs9  pair is recognized as two sources  in GDR3, which
gives  a  parallax of  4.93$\pm$0.03\,mas  (RUWE  1.6)  for A  and  no
parallax for B.

\subsection{HIP 49336 (Quadruple)}

\begin{figure}[ht]
\plotone{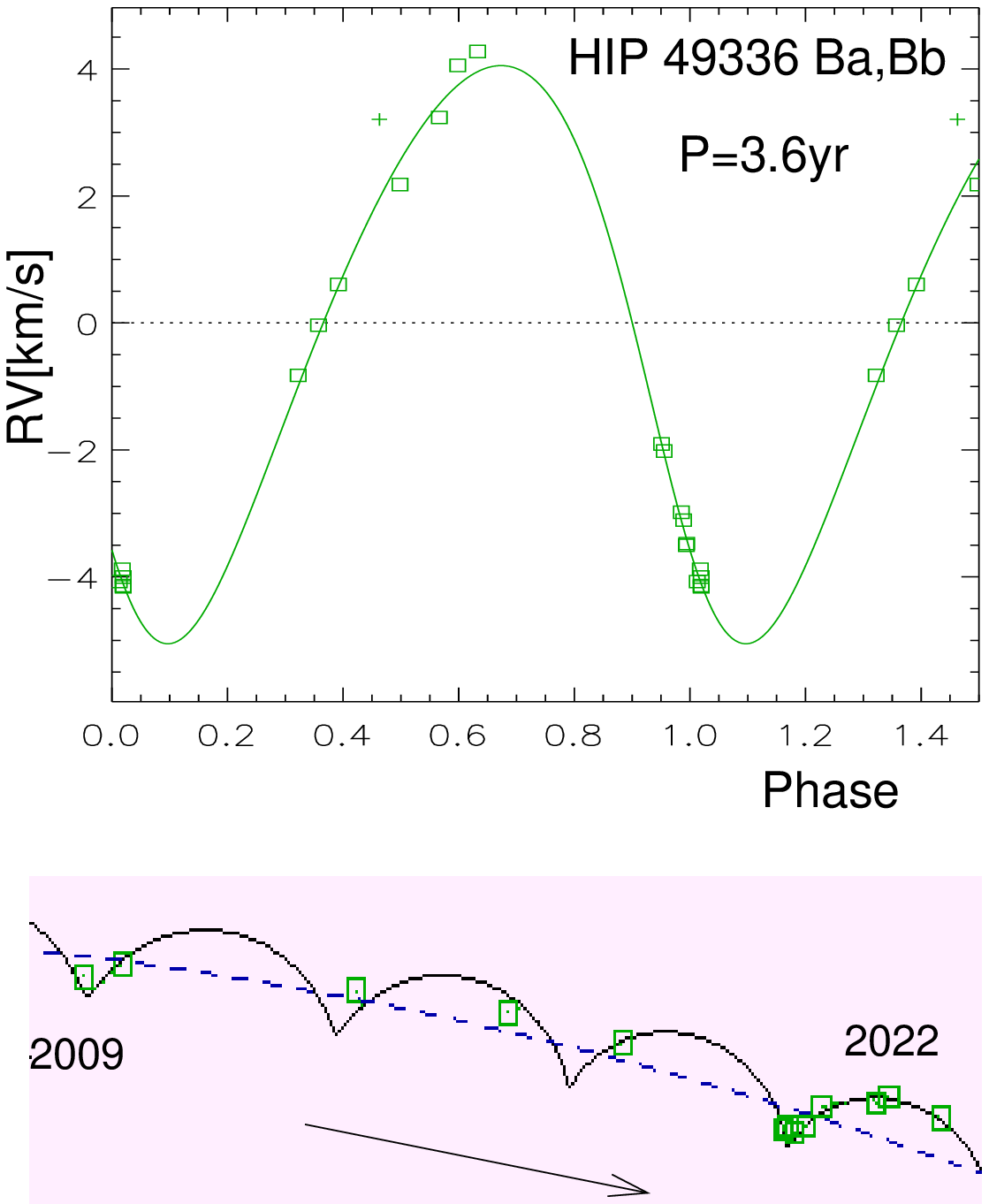}
\caption{The RV  curve of HIP  49336 Ba,Bb  (top, the plus  sign marks
  uncertain measurement) and a fragment of the outer orbit with wobble
  (bottom).  The black solid and blue dashed lines represent the outer
  orbit with and  without wobble, respectively. The  green squares are
  the measured positions of B relative to A.
\label{fig:49336} 
}
\end{figure}

Like the  previous object, this  quadruple system is a  left-over from
the previous work \citep[paper~7,][]{chiron7},  where the 44.5-day orbit
of the  main subsystem  Aa,Ab was established.   The outer  pair I~292
(ADS~7629) has a visual orbit  with $P=380$~yr and $a=0\farcs869$.  It
is  not resolved  by CHIRON,  and the  spectra are  triple-lined.  The
lines of Ba, free  from blending when the lines of Aa  and Ab are well
separated, show a  slow RV variation detected in  paper 7.  Monitoring
with CHIRON at favorable phases of  Aa,Ab has continued for a few more
years (with an interrupt for COVID-19) and now the orbit of Ba,Bb with
a period of 3.6 yr is sufficiently well constrained.

When the  existence of a  long-period subsystem was  established, more
frequent  speckle  observations at  SOAR  were  scheduled in  hope  of
detecting  the  wobble.   Indeed,  as  shown in  the  lower  panel  of
Figure~\ref{fig:49336}, the  apparent motion of A,B  deviates from the
smooth blue line describing the outer  orbit.  The elements of A,B and
Ba,Bb  were  fitted jointly  with  {\tt  ORBIT3} using  both  position
measurements and  RVs. This  helps to better  constrain the  period of
Ba,Bb and  defines the  orientation of its  orbit.  The  RV difference
between A  and B identifies  the correct  ascending node of  the outer
orbit  and the  mutual inclination,  33\degr.  The  small eccentricity
$e_{\rm Ba,Bb}=0.16$  indicates absence of the  Lidov-Kozai cycles, in
agreement with moderate mutual inclination.

The  inclination  of Ba,Bb  determined  from  the  wobble and  the  RV
amplitude lead  to the Bb  mass of 0.50 \msun,  assuming that Ba  is a
1.35 \msun  star. The resulting  mass ratio $q_{\rm Ba,Bb}=  0.37$ and
the semimajor  axis of 31.4\,mas imply  a wobble with an  amplitude of
8.5\,mas, similar to 7.1\,mas found from fitting the A,B positions.


\subsection{HIP 75663  (Quadruple)}

\begin{figure}[ht]
\plotone{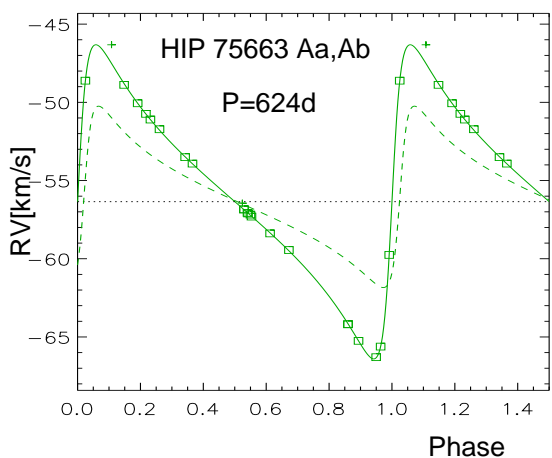}
\caption{The RV curve of HIP 75663  Aa,Ab. Dashed line is the RV curve
  of the GDR3 orbit with corrected $\omega$.
\label{fig:75663} 
}
\end{figure}

Components A and  B of the 9\farcs4 visual  binary STF~1939 (ADS~9640)
are resolved by  the CHIRON 2\farcs7 fiber aperture, so  their RVs are
measured separately.  As established  in paper 4 \citep{chiron4}, star
B is a double-lined twin binary with a period of 22.9 days and $e_{\rm
  Ba,Bb}  = 0.61$,  while  the RV  of  A varies  with  a long  period.
Continued CHIRON monitoring  leads to an orbital period  of 623.8 days
(1.7 yr),  see Figure~\ref{fig:75663}.  The periastron  in 2020.68 was
missed because of the telescope closure to COVID-19, but the following
periastron of  this eccentric ($e_{\rm Aa,Ab}=0.65$)  orbit in 2022.39
has been well covered.  I also used six RVs from \citet{Butler2017} with
an offset of $-56.85$ \kms chosen  to fit the orbit (the published RVs
have arbitrary zero point).

Gaia  DR3 independently determined a  spectro-astrometric orbit of
HIP 75663A with  a period of 626.6758 days and  amplitude $K_1 = 5.80$
\kms.   The general  character of  this orbit  is similar  to the  one
presented here, although the argument of periastron $\omega = 94\fdg5$
is inverted. The dashed line  in Figure~\ref{fig:75663} shows the GDR3
orbit  with $\omega$  corrected by  180\degr.  Fitting  an astrometric
orbit removes the bias of parallax and PM, leading to a good agreement
between parallaxes of A (7.73\,mas) and B (7.78\,mas); the biased
parallax of A in GDR3 is 8.97\,mas with a RUWE of 5.45.  Note that the
short period of Ba,Bb and the equality of its components make its GDR3
astrometry bias-free (RUWE 1.05).

As  noted in  paper  4, star  A  is located  slightly  above the  main
sequence  (estimated  age  $\sim$4  Gyr),  and  the  lithium  line  is
detectable in the spectra of both A  and B.  The Aa mass of 1.47 \msun
estimated from the  standard relation for dwarfs  is only approximate.
The corresponding  minimum mass of Ab  derived from our orbit  is 0.48
\msun.   If the  orbital inclination  of 69\fdg1  measured by  Gaia is
adopted, the mass of Ab becomes 0.52 \msun.  On the other hand, the Ab
mass derived from the GDR3 astrometric  orbit is 0.46 \msun, less than
the minimum  spectroscopic mass, while  the small RV amplitude  in the
GDR3 orbit leads to a minimum mass  of 0.09 \msun.  The actual mass of
Ab should  therefore be close to  0.5 \msun and the  semimajor axis of
the Aa,Ab orbit is 14\,mas.


\subsection{HIP 78163  (Quadruple)}

\begin{figure}[ht]
\plotone{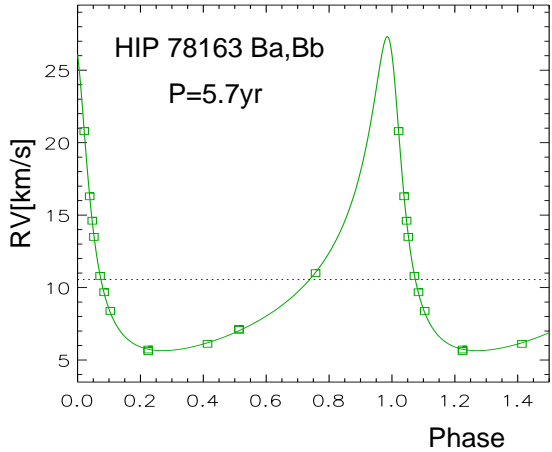}
\caption{The RV curve of HIP 78163  Ba,Bb. 
\label{fig:78163} 
}
\end{figure}

The  2+2 quadruplet  HIP~78163 resembles  the previous  one, but  with
inverted roles  of the components.   The double-lined twin  pair Aa,Ab
with $P=21.8$ days and $e_{\rm Aa,Ab}  = 0.58$ is very similar to star
B in HIP~75663 (22.9 days, $e=0.61$); its orbit has been determined in
paper  4 of  this  series  \citep{chiron4}.  Star  B  of HIP~78163  is
located at  5\farcs9 from A (WG~185 pair in the WDS, estimated  period 7.4
kyr).  The RV of B varies slowly, and, as for the previous object, Gaia
determined an  orbit with  a period  of 1532 days,  this time  only an
astrometric one.  The period found here is longer, 2083 days (5.7 yr).
The Gaia astrometric orbital fit gives  a parallax of 10.65\,mas for B
which agrees much better with  the 10.49\,mas parallax of A (unbiased,
RUWE 0.86).  In contrast, the raw (biased, RUWE 6.22) GDR3 parallax of
B is 11.28\,mas,  while DR2 measured an even  more discrepant parallax
of 13.57\,mas. The duration of the  GDR3 mission is only 34 months, so
a more accurate orbit of Ba,Bb is expected in the future releases.

The spectroscopic  orbit of  Ba,Bb shown in  Figure~\ref{fig:78163} is
eccentric, $e_{\rm Ba,Bb}=0.62$.  The RV maximum is not fully covered,
but the next  periastron is expected only  in 2026.  I use  with a low
weight the RV measured in 2015.5 by Gaia because the CHIRON data cover
only 1690  days.  Adopting a  mass of 0.93  \msun for Ba,  the minimum
mass of  Bb is  0.71 \msun.  Lines  of Bb might  be detectable  in the
spectra, unless  it is  a white  dwarf.  However,  the spectra  can be
partially contaminated by the light of  A, depending on the seeing and
guiding (the separation is only 5\farcs9), so accurate modeling of the
CCFs needed to extract the RVs of Bb is problematic.


\subsection{HIP 79979+79980 (Triple)}

\begin{figure}[ht]
\plotone{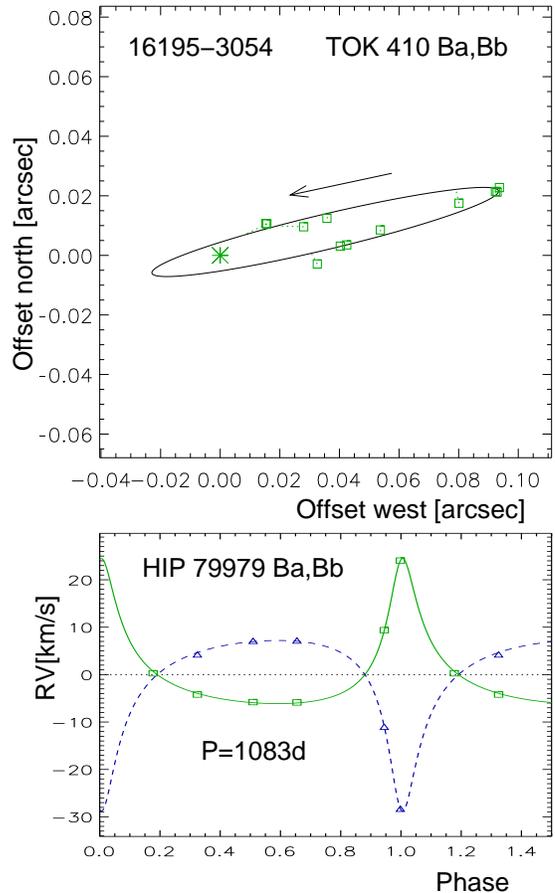}
\caption{Visual orbit and RV curve of HIP 79979 Ba,Bb.
\label{fig:79980} 
}
\end{figure}

The outer 23\farcs4 pair A,B (BSO~12)  has been known since 1837.  Its
brighter component A ($V=5.51$ mag, F5IV) is listed in the bright star
catalog as HR~6077. The fainter ($V=6.82$ mag, F9V) star B has its own
designations HIP~79979  and HD~146835.   The RV  variability of  B was
suspected by  \citet{N04}.  The  RVs of  A and B  were found  equal in
\citep{LCO}, casting  doubt on the  existence of a subsystem,  but the
first CHIRON  spectrum taken in 2017  produced a double CCF.   By that
time, B has been resolved at SOAR as a tight visual pair TOK~410.  The
preliminary orbit  with $P=3$  yr predicted  periastron in  2022.6, as
actually observed (Figure~\ref{fig:79980}).

Using the unbiased  parallax of 22.71 mas measured in  GDR3 for A, the
Ba,Bb  orbit gives  the  mass  sum of  2.11  \msun.  This matches  the
spectroscopic masses of  1.14 and 0.97 \msun  (the inclination $i_{\rm
  Ba,Bb} =  82\fdg9$ is known) and  the absolute magnitudes of  Ba and
Bb. So, despite the modest number of RVs, the orbit of Ba,Bb is
reasonably well defined. 

Stars   A   and   B   have  almost   identical   $V-K$   colors   (see
table~\ref{tab:objects}), but differ  by 1.3 mag in the $V$  band. Star A
is obviously evolved; it is located above the main sequence. In contrast,
star B, despite being a binary, is located on the standard main
sequence.  

\subsection{HIP 103735  (Triple)}

\begin{figure}[ht]
\plotone{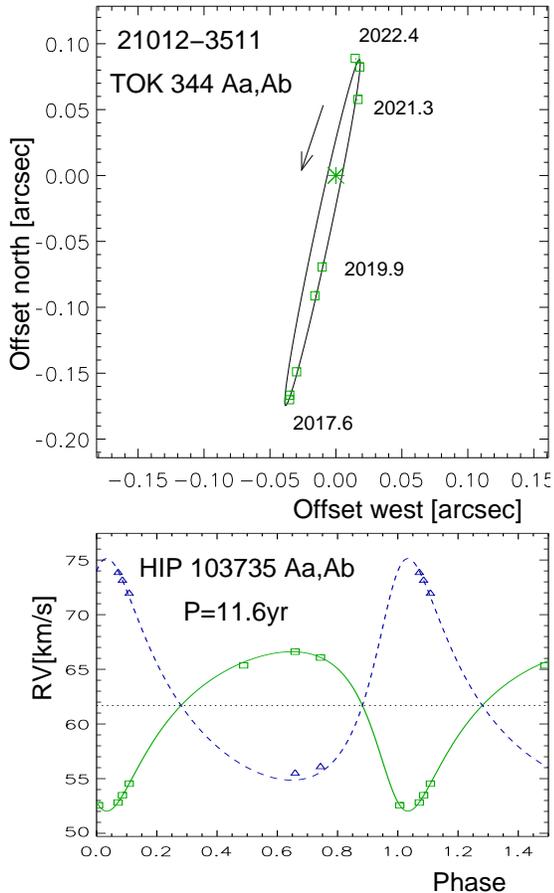}
\caption{Visual orbit and RV curve of HIP 103735 Aa,Ab.
\label{fig:103735} 
}
\end{figure}

The primary  component A  ($V=7.66$ mag, G3V)  of the  wide 186\arcsec
~pair  is a  visual and  spectroscopic  binary. The  secondary star  B
(2MASS J21012669-3509333,  $V=17.14$ mag) is a  white dwarf identified
by \citet{TokLep2012} in the large PM survey and confirmed by Gaia.

Both  \citet{Nidever2002} and  \citet{N04}  noted that  RV  of A  was
variable.   The  first  CHIRON  spectrum taken  in  2017  revealed  an
asymmetric (blended) CCF.  The  same year the 0\farcs14 pair Aa,Ab
has  been resolved  at SOAR  (TOK 344  Aa,Ab).  In  the following  five
years, the pair passed though the periastron: the separation decreased
and increased  again, the CCF  dips separated apart. These  data allow
calculation  of  a  combined  orbit  with  $P=11.6$  yr  presented  in
Figure~\ref{fig:103735}.  One RV  published by \citet{Nidever2002} is
used, it refers to the brighter star Aa.

The combined orbit yields masses of 1.00 and 0.72 \msun for Aa and Ab,
respectively,  and an  orbital parallax  of 21.5\,mas,  in rough
agreement with the  accurate GDR3 parallax of star B,  22.09\,mas. The GDR3
parallax  of A  is  inaccurate and  biased, 23.55$\pm$0.47\,mas.   The
astrometric acceleration  is reflected by  the large RUWE of  15.6, as
well as by the PM anomaly \citep{Brandt2021}.


\subsection{HIP 103814 (Triple)}

\begin{figure}[ht]
\plotone{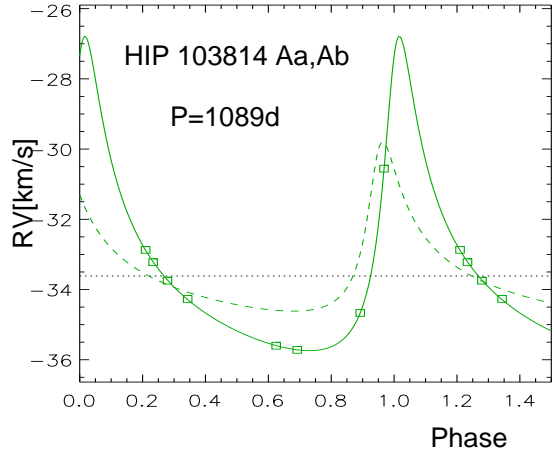}
\caption{RV curve of HIP 103814 Aa,Ab. The GDR3 orbit is traced by the
  dashed line.  
\label{fig:103814} 
}
\end{figure}

The 57\arcsec ~pair of bright stars HIP 103814 (HR~8042, $V=6.64$ mag,
G3IV+K0IV) and  HIP 103839 ($V=6.90$  mag, K0III) has been known  since 1826
(DUN~236 in  the WDS).   B is redder  than A and  brighter in  the $K$
band. This  is a  rare pair composed  of two giants,  and it  does not
belong to the 67-pc sample of solar-type stars.

The  fact  that star  A  is  a  binary  follows from  its  astrometric
acceleration detected by Hipparcos,  RUWE of 23.0 in GDR3, and, possibly,
composite spectrum.  The  eight CHIRON RVs do not  fully constrain the
orbit   shown   in    Figure~\ref{fig:103814}.   However,   the   GDR3
spectro-astrometric orbit  with $P=1119.65$  days confirms  the period
independently.  The shape  of  the  Gaia RV  curve  is similar  (after
correcting   $\omega$  by   180\degr),  although   its  amplitude   is
substantially smaller (2.41  \kms) compared to the  CHIRON orbit (4.48
\kms). 

Assuming that the mass of Aa is 1.78 \msun, the GDR3 astrometric orbit
with an amplitude of 7.7\,mas corresponds to the Ab mass of 0.53 \msun
(the full semimajor  axis is 33.3 mas).  This implies  a early-M dwarf
companion which  contributes negligible  light, so  the spectrum  of A
cannot be composite. The minimum Ab mass derived from the CHIRON orbit
is 0.28  \msun, and  the actual  mass is  0.62 \msun,  considering the
inclination of the  astrometric orbit.  The width and  contrast of the
CCF dip  do not  change with  orbital phase, proving  that Ab  is much
fainter than Aa.


\subsection{HIP 104440 (Triple)}

\begin{figure}[ht]
\plotone{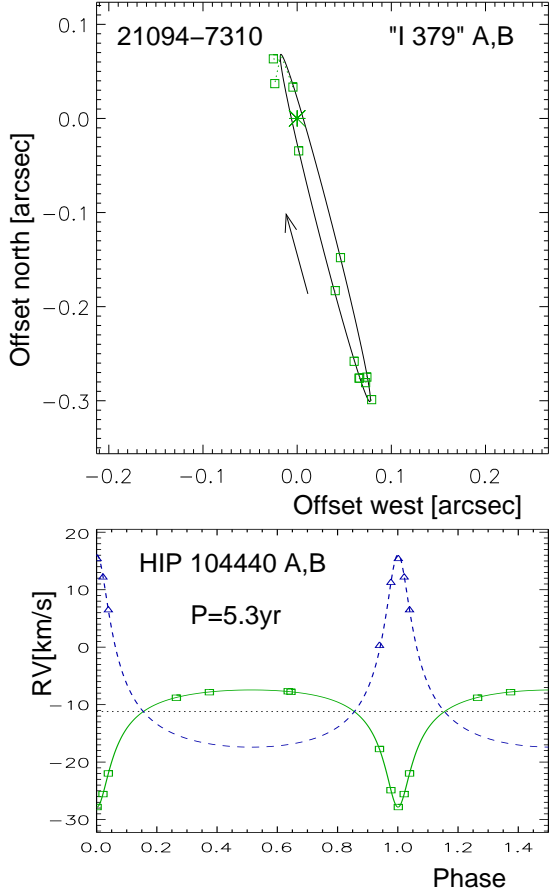}
\caption{Visual orbit and RV curve of HIP 104440 A,B.
\label{fig:104440} 
}
\end{figure}

This is  a resolved visual triple  located at 20\,pc from  the Sun (GJ
818.1).   The outer  6\farcs4  pair  AB,C has  been  known since  1894
(HDO~305).  Star  C is  faint ($V=13.5$  mag) and  red, likely  an M4V
dwarf. AB,C is  in slow retrograde motion with an  estimated period of
1\,kyr. The PM difference between AB and  C is caused by motion in the
outer orbit.

The bright  ($V=5.68$ mag, F9.5V) visual  pair A,B known as  I~379 has
been presumably discovered  by R.~Innes in 1898, although  we know now
that the  separations on the order  of 1\arcsec ~measured by  him were
totally wrong  (this pair is  never wider than 0\farcs3).   Apart from
the three  spurious measurements by  Innes, only W.~Finsen  reported a
resolution of this  pair in 1932 which also does  not match the orbit.
The magnitude difference measured at  SOAR is substantial, $\Delta y =
3.15$ mag,  and such  close pairs  are beyond  the capacity  of visual
observers. In this case, the  WDS name I~379 corresponds to
the   spurious  discovery,   despite  several   ``confirming''  visual
resolutions.

\citet{Goldin2007} published  two possible astrometric orbits  of this
star with  periods of 6.65  and 5.87  yr based on  Hipparcos transits.
The true  period is even shorter,  5.3 yr.  The first  visual orbit of
A,B  which   also  used   the  CHIRON  RVs   has  been   published  in
\citep{Tok2020};  it is  updated here  (Figure~\ref{fig:104440}).  The
pair goes though the periastron in 2022.9, and the previous periastron
in 2017.4 has been also  covered.  The orbit ignores spurious historic
micrometer measurements and  is based entirely on the  SOAR and CHIRON
data.

The absolute  magnitudes of A and  B correspond to the  masses of 1.13
and 0.72 \msun and a dynamic parallax of 49.7\,mas which compares well
with the  GDR3 parallax of star  C, 50.6\,mas; the GRD3  parallax of A,
47.0\,mas, is biased. Masses derived  from the combined orbit are 1.15
and 0.74 \msun, and the orbital parallax is 51.0\,mas.

\subsection{HIP 105879 (Triple)}

\begin{figure}[ht]
\plotone{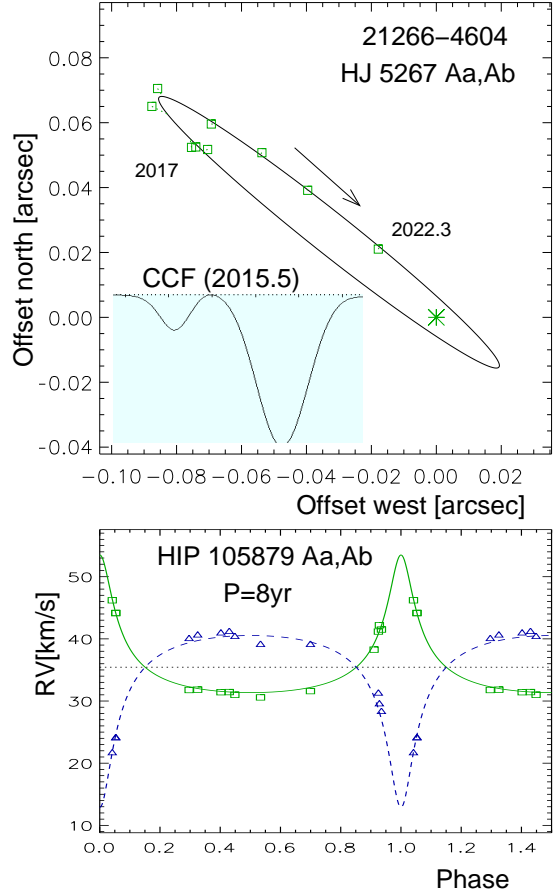}
\caption{Visual orbit and RV curve of HIP 105879 Aa,Ab. The insert
  shows the CCF recorded on JD 24\,57218 when the dips were well
  separated. 
\label{fig:105879} 
}
\end{figure}

This is yet another typical  solar-type triple system composed of wide
and  tight  visual  pairs.   Star  A  ($V=7.18$  mag,  F7V)  has  been
identified  as a  double-lined  binary with  CHIRON  in 2015.5,  first
resolved  at SOAR  in 2017.6,  and designated  in the  WDS as  HJ~5267
Aa,Ab.   The   variable  RV  was  noted   previously  by  \citet{N04},
astrometric  acceleration  was  detected   by  Hipparcos  and  by  its
comparison with  Gaia. The  wide companion D  (CD$-$46~13953, $V=9.96$
mag) at  44\arcsec ~has a  matching PM and  RV.  Its GDR3  parallax of
13.102$\pm$0.014 mas defines accurate distance to the system.  The
companion B,  seen only once  by J.~Herschel  in 1834 at  5\arcsec, is
spurious, and  the companion  C at  238\arcsec ~listed  in the  WDS is
optical.  So, to the best of our knowledge, this is a triple system.

The  combined  orbit  of  Aa,Ab  with $P=8.0$  yr  and  a  substantial
eccentricity    $e_{\rm   Aa,Ab}    =    0.63$    is   presented    in
Figure~\ref{fig:105879}. The  first spectrum has been  taken in 2010.8
using fiber echelle \citep{Tok2015}, and  the 11.9 yr coverage defines
the orbital period quite well.  The pair Aa,Ab goes through periastron
in 2023.2, when it will not  be visible behind the Sun. Unfortunately,
the period is an integer number  of years and in the foreseeable future
all periastrons will  occur during poor visibility  periods.  The pair
was unresolved at SOAR in 2015.74 and in 2022.68 in agreement with the
orbit that predicted small separations on those dates.

The  CCF  dips  are  well  separated  only  near  the  periastron,  as
illustrated in the  Figure. In other phases they are  blended, and the
fits of two overlapping Gaussians are  less reliable. The ratio of the
dip areas  when they are  well separated corresponds to  the magnitude
difference of 2.21 mag, in  agreement with the differential photometry
at SOAR ($\Delta y = 2.14$ mag, $\Delta I = 1.94$ mag). This relatively
large magnitude  difference does not match  the moderate spectroscopic
mass ratio $q_{\rm Aa,Ab} = 0.80$. The  mass ratio and the mass sum of
2.06 \msun derived from the visual elements and the parallax of star D
lead to the  individual masses of 1.14  and 0.92 \msun for  Aa and Ab,
respectively, while the absolute magnitude of Aa corresponds to a mass
of 1.5 \msun  on the main sequence.  In fact, A is  elevated above the
main  sequence  by  $\sim$1  mag,  so  Aa  starts  to  evolve  into  a
subgiant. This  explains the  apparent discrepancy between  mass ratio
and magnitude difference in the inner pair. The spectroscopic mass sum
is  2.2 \msun,  suggesting that  the RV  amplitudes might  be slightly
over-estimated.

Star D has not been resolved by speckle interferometry at SOAR, it has
low  astrometric noise  in Gaia  and  an apparently  constant RV  that
matches the RV of A. So, it is unlikely that D has close companions.

\subsection{HIP 109443 (Triple)}

\begin{figure}[ht]
\plotone{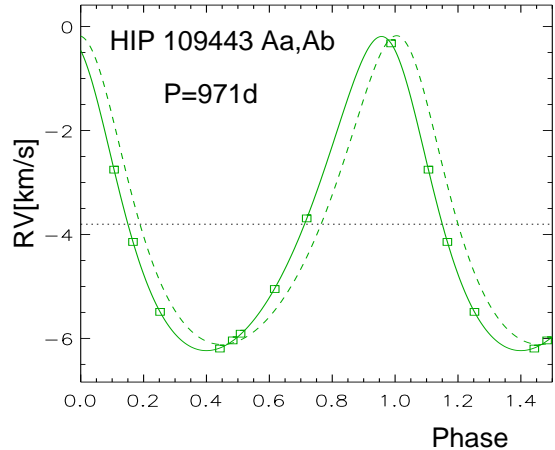}
\caption{RV curve of HIP 109443 Aa,Ab. The dashed line is the Gaia
  spectroscopic orbit. 
\label{fig:109443} 
}
\end{figure}

The  bright  solar-type star  HIP~109443  ($V=7.63$  mag, F8V)  is  an
astrometric  binary  detected  by  Hipparcos  and  confirmed  both  by
\citet{Brandt2021}  and by  a  RUWE  of 10.3  in  GDR3.   A survey  of
astrometric  binaries with  the NICI  AO instrument  detected a  faint
companion  B at  1\farcs4  separation \cite[][TOK~216]{Tok2012}.   The
estimated  period  of   $\sim$700  yr  makes  it   unlikely  that  the
acceleration and variable RV \citep{N04} are caused by this companion.

Nine  CHIRON  spectra  show  the RV  variability,  and  the  GDR3
spectroscopic     orbit    matches     these     RVs    quite     well
(Figure~\ref{fig:109443}). The fit of 6 elements to 9 RVs is almost
perfect, leaving rms residuals of only 0.008 \kms. 

The  minimum mass  of  Ab is  0.18  \msun if  the mass  of  Aa is  1.3
\msun. The  large RUWE  indicates clear  detection of  the astrometric
signal, but, for  some reason, GDR3 determined  only the spectroscopic
orbit,   leaving   the  inclination   and   the   true  mass   of   Ab
unconstrained. The mass of B is about 0.56 \msun, as inferred from its
$K$-band luminosity.  It is detected  by Gaia at  1\farcs4129 ($\Delta
G_{\rm  A,B} =  4.85$ mag),  showing  little motion  since 2011.   The
parallax of B  measured by GDR3, 15.58\,mas, is close  to the parallax
of  A in  GDR2  (15.33\,mas),  but GDR3  gives  a  biased parallax  of
12.74\,mas for A.  The bias will be removed when the astrometric orbit
of Aa,Ab is determined by Gaia.

\subsection{HIP 117666 (Quadruple)}


\begin{figure}[ht]
\plotone{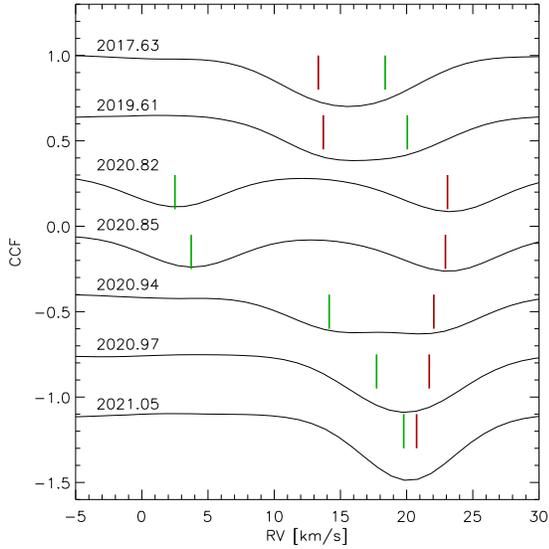}
\caption{CCFs of the first 7 CHIRON  spectra of HIP 117666. The curves
  are displaced  vertically by  0.35 to avoid  overlap. The  dates are
  indicated.  Thick red and green lines mark the  RVs of Aa  and Ba,
  respectively, according to the orbits.
\label{fig:117666ccf} 
}
\end{figure}

\begin{figure}[ht]
\plotone{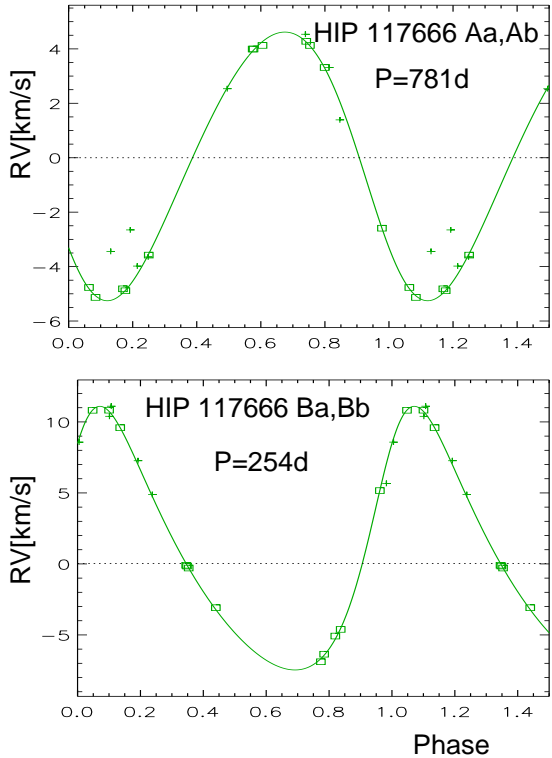}
\caption{RV  curves  of   the  subsystems  Aa,Ab  and   Ba,Bb  in  HIP
  117666. Squares  mark the RVs  derived from double CCFs,  plus signs
  correspond  to  blended   CCFs.   Motion  in  the   outer  orbit  is
  subtracted.
\label{fig:117666} 
}
\end{figure}

\begin{figure}[ht]
\plotone{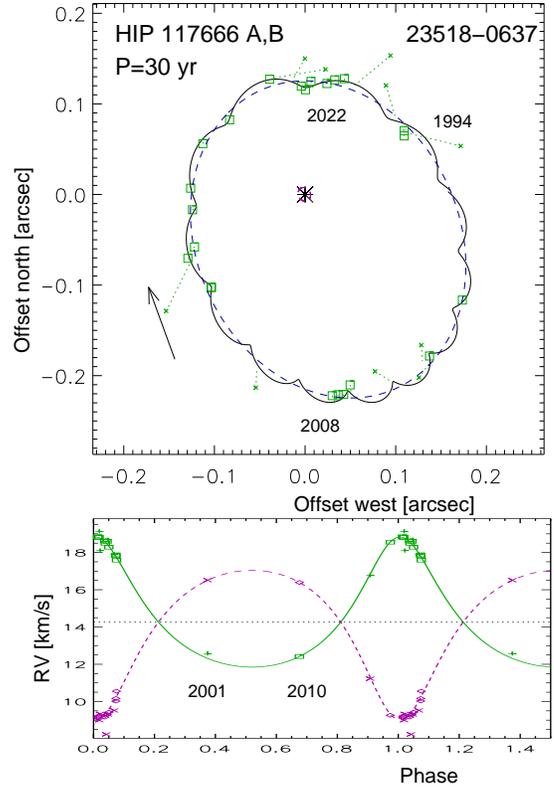}
\caption{Outer orbit  of HIP 117666  A,B on the  sky (top) and  the RV
  curve where motion in the inner subsystems is subtracted (bottom). 
\label{fig:117666out} 
}
\end{figure}

This 9th  magnitude star  (HD 2236888,  G5V) appeared  to be  a normal
tight  visual binary,  first resolved  by R.~Aitken  in 1913  (A~2700,
ADS~17052, WDS J23518$-$0637).  The visual  orbit with a 30 yr period,
last updated  by \citet{Docobo2009}, is  very well constrained.  It is
fully covered by accurate speckle measurements (the first one in 1985)
and is rated grade 2 in the orbit catalog.

The star  attracted attention as an  X-ray source and for  this reason
four  high-resolution  spectra  were taken  by  \citet{Frasca2018}  in
2001. Double lines were detected  in one spectrum, suggesting presence
of a subsystem.  Double lines were  also seen in the spectrum taken by
\citet{Tok2015}  in  2010.   This  prompted  further  monitoring  with
CHIRON. It  was not clear  from the  outset whether double  lines were
produced by motion in the visual  orbit or by an inner subsystem.  The
first  CHIRON spectrum  taken in  2017.6 looked  single-lined, but  in
2019.6 the lines  were double again, so I  started regular monitoring.
The first  seven CHIRON  CCFs are shown  in Figure~\ref{fig:117666ccf},
illustrating blending of two components with independent RV variation.
The RVs  of both CCF  dips vary faster  than prescribed by  the visual
orbit, so this system is actually a 2+2 quadruple.

Five RVs  of HIP 117666 averaging  at $\sim -$1 \kms  were reported by
\citet{TS2002}.   They are  not compatible  with the  orbits presented
here. It turns out that these RVs refer to another 9th magnitude star,
HD~223695 (F6IV), which is located at  1$'$ east of HIP 117666 and has
an RV of  $-2.3$ \kms according to Simbad. Somehow,  the visual binary
was misidentified,  and the spectral type  of F6V quoted in  the above
paper  (instead of  G5V)  confirms  this suspicion.   One  of the  RVs
reported by Frasca et al.  (JD 24\,52209.43) also apparently refers to
HD~223695.  Proximity  of two stars  on the  sky means that  the X-ray
source can be actually associated with HD~223695 (with a fast rotation
of 22.0 \kms according to Frasca  et al.), rather than with the slowly
rotating components of the HIP  117666 system. The Galactic velocity of
HIP~117666,  $U,V,W =  (-23.8,-10.1,-21.3)$ \kms,  does not  match any
known kinematic  group, and the  lithium line  in its spectrum  is not
detectable, so this star is not young.

Deciphering  the two  inner orbits  from the  often-blended CCFs  with
similar dips,  superposed on the  slow RV  variation due to  the outer
orbit,  was a  challenging  task, addressed  iteratively.  First,  the
faster variation  of the  smaller dip  Ba was matched  to a  period of
$\sim$250 days,  then the RVs of  Aa yielded a  period of 2 yr  with a
different systemic velocity.  A model was constructed to represent the
blended  dips  by  the  outer  and two  inner  orbits  using  the  dip
amplitudes and widths measured in the well-resolved phases. Comparison
of all  CCFs to this  model inspired  confidence in the  adopted inner
periods.

The available tool, {\tt ORBIT3},  does not allow for simultaneous fit
of all three orbits. After several iterations on the orbits of A,B and
Aa,Ab, the RVs of Ba were corrected  for the motion in the outer orbit
and fitted  as a single-lined  binary.  Then  the motion of  Ba,Bb was
subtracted from the RVs of Ba, and  a joint fit of the remaining inner
orbit  Aa,Ab and  the outer  orbit A,B  was done  using {\tt  ORBIT3},
including the speckle measurements.  The astrometric elements of Aa,Ab
(wobble)  were also  determined. The  rms residuals  of accurate  SOAR
speckle positions without  wobble, 6\,mas, are reduced  to 2\,mas when
the wobble is fitted. The smaller wobble caused by the Ba,Bb motion is
left unmodeled. In the orbital fit, I used the RVs measured in 2002 by
Frasca et al.  with a small weight and a correction of +1 \kms.

Speckle interferometry  at SOAR  establishes the  magnitude difference
$\Delta  m_{\rm A,B}  = 0.14$  mag (WDS quotes  0.2 mag),  so the
individual  $V$  magnitudes  of  A  and  B  are  9.41  and  9.55  mag,
respectively.  The Gaia DR2 parallax of 13.4$\pm$1.3 mas is inaccurate
and  potentially  biased, GDR3  gives  no  parallax,  so I  adopt  the
dynamical  parallax of  13.2\,mas that  follows from  the good-quality
outer  orbit and  the estimated  masses. Neglecting  the light  of the
secondaries Ab and Bb, the masses of Aa and Ba matching their absolute
magnitudes  are 0.98  and 0.97  \msun;  they agree  with the  combined
spectral type  G5V and the $V-K$  color.  The mass of  Ab deduced from
the  RV amplitude  and  inclination  (see below)  is  0.45 \msun,  the
minimum  mass  of  Bb  is  0.31 \msun.   Small  masses  of  the  inner
secondaries explain  why their lines  are not visible in  the spectra.
The estimated system mass is therefore 2.70 \msun.

The estimated masses,  periods, and parallax yield  the semimajor axes
of the  inner orbits, 25.1  and 11.2 mas.   The ratio of  the measured
photo-center  amplitude  of Aa,Ab,  7.9\,mas,  to  the semimajor  axis
equals  the wobble  factor  $f_{\rm Aa,Ab}  = q_{\rm  Aa,Ab}/(1+q_{\rm
  Aa,Ab}) =  0.31$, hence the  mass ratio is $q_{\rm  Aa,Ab}=0.46$ and
the estimated  mass of  Ab is  0.45 \msun.   The astrometric  orbit of
Aa,Ab  yields  a  loosely  constrained inclination  $i_{\rm  Aa,Ab}  =
141\degr  \pm 13\degr$.   The inner  inclination of  $i_{\rm Aa,Ab}  =
143\degr$ matches  the Ab mass estimated  from the wobble, so  I fixed
this  value.  The  minimum mass  of Bb  leads to  an estimated  wobble
amplitude  of  2\,mas   for  Ba.   This  unmodeled   (so  far)  wobble
contributes to the residuals of accurate speckle positions.

As  noted, the  outer orbit  (Figure~\ref{fig:117666out}) is  very well
constrained.   However,  the  fitted  inclination of  $i_{\rm  A,B}  =
149\fdg6 \pm 1\fdg5$ and the RV  amplitudes of 3.42 and 4.15 \kms lead
to the  outer mass sum  of 3.25  \msun, substantially larger  than the
estimated 2.7 \msun  and in disagreement with  the absolute magnitudes
and spectral  type. This  discrepancy is removed  by fixing  the outer
inclination  to  a  slightly  smaller  value  of  147\fdg5.   So,  the
inclinations  of  both Aa,Ab  and  A,B  are fine-tuned  (within  limits
allowed  by  the  data)  to   reach  consistency  between  all  orbital
parameters and the estimated masses. The nodal angles $\Omega$ in both
orbits are similar,  hence Aa,Ab and A,B have well  aligned orbits. If
the orbit of Ba,Bb  were also coplanar with A,B, the  mass of Bb would
be 0.55  \msun, making B  more massive  than A.  This  contradicts the
measured  outer mass  ratio  $q_{\rm A,B}  =  0.82$.  The  inclination
$i_{\rm  Ba,Bb}$ could  be close  to 110\degr  ~in order  to avoid  the
Lidov-Kozai oscillations (the eccentricity of Ba,Bb is only 0.27).

The  architecture  of  this  remarkable  quadruple  and  its  internal
dynamics deserve further investigation.  The period ratio $P_{\rm A,B}
/P_{\rm Aa,Ab} =  14.05$ is small, so the  orbits interact dynamically
and  the motion  is  more  complex than  the  simple superposition  of
Keplerian orbits fitted here; our orbits represent osculating elements
in  the current  epoch.  Furthermore,  the ratio  of two  inner periods
$P_{\rm  Aa,Ab} /P_{\rm  Ba,Bb}  \approx  3$ is  close  to an  integer
number.  The  two inner orbits  could be  trapped into a  1:14:42 mean
motion resonance with the outer orbit.


\section{Summary}
\label{sec:sum}

The original  goal of this  project has been  inspired by the  lack of
known  orbital elements  for inner  subsystems revealed  previously by
variable  RV or  astrometric  acceleration.  Periods  and mass  ratios
inferred  from the  orbits are  necessary for  a statistical  study of
nearby hierarchies.   However, our RV  monitoring has shown  that some
hierarchies, believed  to be  triple, are in  fact 2+2  quadruples, as
both visual components have variable  RVs.  Orbits of the short-period
subsystems  in  these  quadruples  have been  quickly  determined  and
reported in  the first  papers of this  series.  Observation  of other
components with  slow RV variation  continued for several   years,
eventually  yielding their  spectroscopic orbits  as well.   Four such
``returning clients'' from the previous papers are presented here: HIP
41171, 49336, 75663, and 78163.  In the first two, the outer pairs are
sub-arcsecond,  so  the spectra  of  all  visible stars  are  blended.
Recovering  RVs  of slowly  varying  subsystems  from blended  spectra
dominated  by  fast  RV  variation of  other  pairs  required  careful
planning of observations at phases where the blending is minimized.

HIP  117666  is  another  tricky  quadruple studied  here.   It  is  a
well-known visual  binary where  existence of  an inner  subsystem has
been suggested  but remained unproven. Unexpectedly,  our observations
discovered that  both visual  components contain  subsystems.  Patient
accumulation of blended spectra was  needed to decipher the underlying
orbits, including the  outer one. A similar case (HIP  12548) has been
presented in paper 8.

The perturbing effects of binaries  on the astrometry were anticipated
in the design of  the Gaia mission, and GDR3 delivered  a large set of
orbits  which reduce  biases in  the astrometric  solutions.  However,
complexity of  some triple  and quadruple hierarchies  precludes their
modeling  by  the  Gaia  pipeline.   Only  dedicated  monitoring  with
adequate cadence,  time coverage,  and spectral resolution  can reveal
the true nature of systems like HIP 12548 and 117666.  Even relatively
easy cases  of spectroscopic subsystems in  well-resolved visual pairs
(HIP 75663A,  79163B, and 103814A)  demonstrate that some  Gaia orbits
can be  inaccurate or even  wrong. Gaia is an  outstanding astrometric
facility which has not been designed to deal with complex hierarchies.

Long orbital  periods of  subsystems studied  here favor  their direct
resolution.    Six   combined  orbits   based   on   CHIRON  RVs   and
speckle-interferometric measurements at SOAR were determined.  Another
two are added  from modeling wobble in the motion  of their relatively
close   outer  pairs.   Spatial  resolution   (or  wobble   detection)
complements spectroscopic  orbits, allowing  to measure masses  and to
deduce orbit  orientation.  The  latter is particularly  valuable when
orientation  of  the  outer  orbit is  also  known.   Architecture  of
hierarchical systems  (mutual orbit  orientation, periods  and masses)
contains information on their still debated origin.

\begin{acknowledgments} 

I thank operators of the 1.5-m telescope for executing observations of
this  program  and  the   SMARTS  team  for  scheduling  and  pipeline
processing.

The research was funded by the NSF's NOIRLab.
This work  used the  SIMBAD service operated  by Centre  des Donn\'ees
Stellaires  (Strasbourg, France),  bibliographic  references from  the
Astrophysics Data  System maintained  by SAO/NASA, and  the Washington
Double  Star Catalog  maintained  at USNO.  
This work  has made use of  data from the European  Space Agency (ESA)
mission {\it  Gaia} (\url{https://www.cosmos.esa.int/gaia}), processed
by  the {\it  Gaia}  Data Processing  and  Analysis Consortium  (DPAC,
\url{https://www.cosmos.esa.int/web/gaia/dpac/consortium}).     Funding
for the DPAC has been provided by national institutions, in particular
the  institutions   participating  in  the   {\it  Gaia}  Multilateral
Agreement. This research has made use of the services of the ESO
Science Archive Facility.

\end{acknowledgments} 

\facility{CTIO:1.5m, SOAR, Gaia}






\begin{thebibliography}{}
\expandafter\ifx\csname natexlab\endcsname\relax\def\natexlab#1{#1}\fi
\providecommand{\url}[1]{\href{#1}{#1}}
\providecommand{\dodoi}[1]{doi:~\href{http://doi.org/#1}{\nolinkurl{#1}}}
\providecommand{\doeprint}[1]{\href{http://ascl.net/#1}{\nolinkurl{http://ascl.net/#1}}}
\providecommand{\doarXiv}[1]{\href{https://arxiv.org/abs/#1}{\nolinkurl{https://arxiv.org/abs/#1}}}

\bibitem[{{Brandt}(2021)}]{Brandt2021}
{Brandt}, T.~D. 2021, \apjs, 254, 42, \dodoi{10.3847/1538-4365/abf93c}

\bibitem[{{Butler} {et~al.}(2017){Butler}, {Vogt}, {Laughlin}, {Burt},
  {Rivera}, {Tuomi}, {Teske}, {Arriagada}, {Diaz}, {Holden}, \&
  {Keiser}}]{Butler2017}
{Butler}, R.~P., {Vogt}, S.~S., {Laughlin}, G., {et~al.} 2017, \aj, 153, 208,
  \dodoi{10.3847/1538-3881/aa66ca}

\bibitem[{{Docobo} \& {Ling}(2009)}]{Docobo2009}
{Docobo}, J.~A., \& {Ling}, J.~F. 2009, \aj, 138, 1159,
  \dodoi{10.1088/0004-6256/138/4/1159}

\bibitem[{{Frasca} {et~al.}(2018){Frasca}, {Guillout}, {Klutsch}, {Ferrero},
  {Marilli}, {Biazzo}, {Gandolfi}, \& {Montes}}]{Frasca2018}
{Frasca}, A., {Guillout}, P., {Klutsch}, A., {et~al.} 2018, \aap, 612, A96,
  \dodoi{10.1051/0004-6361/201732028}

\bibitem[{{Gaia Collaboration} {et~al.}(2021){Gaia Collaboration}, {Brown},
  {Vallenari}, {Prusti}, {de Bruijne}, {Babusiaux}, {Biermann}, {Creevey},
  {Evans}, {Eyer}, {Hutton}, {Jansen}, {Jordi}, {Klioner}, {Lammers},
  {Lindegren}, {Luri}, {Mignard}, {Panem}, {Pourbaix}, {Randich}, {Sartoretti},
  {Soubiran}, {Walton}, {Arenou}, {Bailer-Jones}, {Bastian}, {Cropper},
  {Drimmel}, {Katz}, {Lattanzi}, {van Leeuwen}, {Bakker}, {Cacciari},
  {Casta{\~n}eda}, {De Angeli}, {Ducourant}, {Fabricius}, {Fouesneau},
  {Fr{\'e}mat}, {Guerra}, {Guerrier}, {Guiraud}, {Jean-Antoine Piccolo},
  {Masana}, {Messineo}, {Mowlavi}, {Nicolas}, {Nienartowicz}, {Pailler},
  {Panuzzo}, {Riclet}, {Roux}, {Seabroke}, {Sordo}, {Tanga}, {Th{\'e}venin},
  {Gracia-Abril}, {Portell}, {Teyssier}, {Altmann}, {Andrae}, {Bellas-Velidis},
  {Benson}, {Berthier}, {Blomme}, {Brugaletta}, {Burgess}, {Busso}, {Carry},
  {Cellino}, {Cheek}, {Clementini}, {Damerdji}, {Davidson}, {Delchambre},
  {Dell'Oro}, {Fern{\'a}ndez-Hern{\'a}ndez}, {Galluccio}, {Garc{\'\i}a-Lario},
  {Garcia-Reinaldos}, {Gonz{\'a}lez-N{\'u}{\~n}ez}, {Gosset}, {Haigron},
  {Halbwachs}, {Hambly}, {Harrison}, {Hatzidimitriou}, {Heiter},
  {Hern{\'a}ndez}, {Hestroffer}, {Hodgkin}, {Holl}, {Jan{\ss}en}, {Jevardat de
  Fombelle}, {Jordan}, {Krone-Martins}, {Lanzafame}, {L{\"o}ffler}, {Lorca},
  {Manteiga}, {Marchal}, {Marrese}, {Moitinho}, {Mora}, {Muinonen}, {Osborne},
  {Pancino}, {Pauwels}, {Petit}, {Recio-Blanco}, {Richards}, {Riello},
  {Rimoldini}, {Robin}, {Roegiers}, {Rybizki}, {Sarro}, {Siopis}, {Smith},
  {Sozzetti}, {Ulla}, {Utrilla}, {van Leeuwen}, {van Reeven}, {Abbas}, {Abreu
  Aramburu}, {Accart}, {Aerts}, {Aguado}, {Ajaj}, {Altavilla}, {{\'A}lvarez},
  {{\'A}lvarez Cid-Fuentes}, {Alves}, {Anderson}, {Anglada Varela}, {Antoja},
  {Audard}, {Baines}, {Baker}, {Balaguer-N{\'u}{\~n}ez}, {Balbinot}, {Balog},
  {Barache}, {Barbato}, {Barros}, {Barstow}, {Bartolom{\'e}}, {Bassilana},
  {Bauchet}, {Baudesson-Stella}, {Becciani}, {Bellazzini}, {Bernet}, {Bertone},
  {Bianchi}, {Blanco-Cuaresma}, {Boch}, {Bombrun}, {Bossini}, {Bouquillon},
  {Bragaglia}, {Bramante}, {Breedt}, {Bressan}, {Brouillet}, {Bucciarelli},
  {Burlacu}, {Busonero}, {Butkevich}, {Buzzi}, {Caffau}, {Cancelliere},
  {C{\'a}novas}, {Cantat-Gaudin}, {Carballo}, {Carlucci}, {Carnerero},
  {Carrasco}, {Casamiquela}, {Castellani}, {Castro-Ginard}, {Castro Sampol},
  {Chaoul}, {Charlot}, {Chemin}, {Chiavassa}, {Cioni}, {Comoretto}, {Cooper},
  {Cornez}, {Cowell}, {Crifo}, {Crosta}, {Crowley}, {Dafonte}, {Dapergolas},
  {David}, {David}, {de Laverny}, {De Luise}, {De March}, {De Ridder}, {de
  Souza}, {de Teodoro}, {de Torres}, {del Peloso}, {del Pozo}, {Delbo},
  {Delgado}, {Delgado}, {Delisle}, {Di Matteo}, {Diakite}, {Diener},
  {Distefano}, {Dolding}, {Eappachen}, {Edvardsson}, {Enke}, {Esquej}, {Fabre},
  {Fabrizio}, {Faigler}, {Fedorets}, {Fernique}, {Fienga}, {Figueras},
  {Fouron}, {Fragkoudi}, {Fraile}, {Franke}, {Gai}, {Garabato},
  {Garcia-Gutierrez}, {Garc{\'\i}a-Torres}, {Garofalo}, {Gavras}, {Gerlach},
  {Geyer}, {Giacobbe}, {Gilmore}, {Girona}, {Giuffrida}, {Gomel}, {Gomez},
  {Gonzalez-Santamaria}, {Gonz{\'a}lez-Vidal}, {Granvik},
  {Guti{\'e}rrez-S{\'a}nchez}, {Guy}, {Hauser}, {Haywood}, {Helmi}, {Hidalgo},
  {Hilger}, {H{\l}adczuk}, {Hobbs}, {Holland}, {Huckle}, {Jasniewicz},
  {Jonker}, {Juaristi Campillo}, {Julbe}, {Karbevska}, {Kervella}, {Khanna},
  {Kochoska}, {Kontizas}, {Kordopatis}, {Korn}, {Kostrzewa-Rutkowska},
  {Kruszy{\'n}ska}, {Lambert}, {Lanza}, {Lasne}, {Le Campion}, {Le Fustec},
  {Lebreton}, {Lebzelter}, {Leccia}, {Leclerc}, {Lecoeur-Taibi}, {Liao},
  {Licata}, {Lindstr{\o}m}, {Lister}, {Livanou}, {Lobel}, {Madrero Pardo},
  {Managau}, {Mann}, {Marchant}, {Marconi}, {Marcos Santos}, {Marinoni},
  {Marocco}, {Marshall}, {Martin Polo}, {Mart{\'\i}n-Fleitas}, {Masip},
  {Massari}, {Mastrobuono-Battisti}, {Mazeh}, {McMillan}, {Messina},
  {Michalik}, {Millar}, {Mints}, {Molina}, {Molinaro}, {Moln{\'a}r},
  {Montegriffo}, {Mor}, {Morbidelli}, {Morel}, {Morris}, {Mulone}, {Munoz},
  {Muraveva}, {Murphy}, {Musella}, {Noval}, {Ord{\'e}novic}, {Orr{\`u}},
  {Osinde}, {Pagani}, {Pagano}, {Palaversa}, {Palicio}, {Panahi}, {Pawlak},
  {Pe{\~n}alosa Esteller}, {Penttil{\"a}}, {Piersimoni}, {Pineau}, {Plachy},
  {Plum}, {Poggio}, {Poretti}, {Poujoulet}, {Pr{\v{s}}a}, {Pulone}, {Racero},
  {Ragaini}, {Rainer}, {Raiteri}, {Rambaux}, {Ramos}, {Ramos-Lerate}, {Re
  Fiorentin}, {Regibo}, {Reyl{\'e}}, {Ripepi}, {Riva}, {Rixon}, {Robichon},
  {Robin}, {Roelens}, {Rohrbasser}, {Romero-G{\'o}mez}, {Rowell}, {Royer},
  {Rybicki}, {Sadowski}, {Sagrist{\`a} Sell{\'e}s}, {Sahlmann}, {Salgado},
  {Salguero}, {Samaras}, {Sanchez Gimenez}, {Sanna}, {Santove{\~n}a},
  {Sarasso}, {Schultheis}, {Sciacca}, {Segol}, {Segovia}, {S{\'e}gransan},
  {Semeux}, {Shahaf}, {Siddiqui}, {Siebert}, {Siltala}, {Slezak}, {Smart},
  {Solano}, {Solitro}, {Souami}, {Souchay}, {Spagna}, {Spoto}, {Steele},
  {Steidelm{\"u}ller}, {Stephenson}, {S{\"u}veges}, {Szabados}, {Szegedi-Elek},
  {Taris}, {Tauran}, {Taylor}, {Teixeira}, {Thuillot}, {Tonello}, {Torra},
  {Torra}, {Turon}, {Unger}, {Vaillant}, {van Dillen}, {Vanel}, {Vecchiato},
  {Viala}, {Vicente}, {Voutsinas}, {Weiler}, {Wevers}, {Wyrzykowski}, {Yoldas},
  {Yvard}, {Zhao}, {Zorec}, {Zucker}, {Zurbach}, \& {Zwitter}}]{gaia3}
{Gaia Collaboration}, {Brown}, A.~G.~A., {Vallenari}, A., {et~al.} 2021, \aap,
  649, A1, \dodoi{10.1051/0004-6361/202039657}

\bibitem[{{Gaia Collaboration} {et~al.}(2022){Gaia Collaboration}, {Arenou},
  {Babusiaux}, {Barstow}, {Faigler}, {Jorissen}, {Kervella}, {Mazeh},
  {Mowlavi}, {Panuzzo}, {Sahlmann}, {Shahaf}, {Sozzetti}, {Bauchet},
  {Damerdji}, {Gavras}, {Giacobbe}, {Gosset}, {Halbwachs}, {Holl}, {Lattanzi},
  {Leclerc}, {Morel}, {Pourbaix}, {Re Fiorentin}, {Sadowski}, {S{\'e}gransan},
  {Siopis}, {Teyssier}, {Zwitter}, {Planquart}, {Brown}, {Vallenari}, {Prusti},
  {de Bruijne}, {Biermann}, {Creevey}, {Ducourant}, {Evans}, {Eyer}, {Guerra},
  {Hutton}, {Jordi}, {Klioner}, {Lammers}, {Lindegren}, {Luri}, {Mignard},
  {Panem}, {Randich}, {Sartoretti}, {Soubiran}, {Tanga}, {Walton},
  {Bailer-Jones}, {Bastian}, {Drimmel}, {Jansen}, {Katz}, {van Leeuwen},
  {Bakker}, {Cacciari}, {Casta{\~n}eda}, {De Angeli}, {Fabricius}, {Fouesneau},
  {Fr{\'e}mat}, {Galluccio}, {Guerrier}, {Heiter}, {Masana}, {Messineo},
  {Nicolas}, {Nienartowicz}, {Pailler}, {Riclet}, {Roux}, {Seabroke}, {Sordo},
  {Th{\'e}venin}, {Gracia-Abril}, {Portell}, {Altmann}, {Andrae}, {Audard},
  {Bellas-Velidis}, {Benson}, {Berthier}, {Blomme}, {Burgess}, {Busonero},
  {Busso}, {C{\'a}novas}, {Carry}, {Cellino}, {Cheek}, {Clementini},
  {Davidson}, {de Teodoro}, {Nu{\~n}ez Campos}, {Delchambre}, {Dell'Oro},
  {Esquej}, {Fern{\'a}ndez-Hern{\'a}ndez}, {Fraile}, {Garabato},
  {Garc{\'\i}a-Lario}, {Haigron}, {Hambly}, {Harrison}, {Hern{\'a}ndez},
  {Hestroffer}, {Hodgkin}, {Jan{\ss}en}, {Jevardat de Fombelle}, {Jordan},
  {Krone-Martins}, {Lanzafame}, {L{\"o}ffler}, {Marchal}, {Marrese},
  {Moitinho}, {Muinonen}, {Osborne}, {Pancino}, {Pauwels}, {Recio-Blanco},
  {Reyl{\'e}}, {Riello}, {Rimoldini}, {Roegiers}, {Rybizki}, {Sarro}, {Smith},
  {Utrilla}, {van Leeuwen}, {Abbas}, {{\'A}brah{\'a}m}, {Abreu Aramburu},
  {Aerts}, {Aguado}, {Ajaj}, {Aldea-Montero}, {Altavilla}, {{\'A}lvarez},
  {Alves}, {Anders}, {Anderson}, {Anglada Varela}, {Antoja}, {Baines}, {Baker},
  {Balaguer-N{\'u}{\~n}ez}, {Balbinot}, {Balog}, {Barache}, {Barbato},
  {Barros}, {Bartolom{\'e}}, {Bassilana}, {Becciani}, {Bellazzini},
  {Berihuete}, {Bernet}, {Bertone}, {Bianchi}, {Binnenfeld}, {Blanco-Cuaresma},
  {Blazere}, {Boch}, {Bombrun}, {Bossini}, {Bouquillon}, {Bragaglia},
  {Bramante}, {Breedt}, {Bressan}, {Brouillet}, {Brugaletta}, {Bucciarelli},
  {Burlacu}, {Butkevich}, {Buzzi}, {Caffau}, {Cancelliere}, {Cantat-Gaudin},
  {Carballo}, {Carlucci}, {Carnerero}, {Carrasco}, {Casamiquela}, {Castellani},
  {Castro-Ginard}, {Chaoul}, {Charlot}, {Chemin}, {Chiaramida}, {Chiavassa},
  {Chornay}, {Comoretto}, {Contursi}, {Cooper}, {Cornez}, {Cowell}, {Crifo},
  {Cropper}, {Crosta}, {Crowley}, {Dafonte}, {Dapergolas}, {David}, {de
  Laverny}, {De Luise}, {De March}, {De Ridder}, {de Souza}, {de Torres}, {del
  Peloso}, {del Pozo}, {Delbo}, {Delgado}, {Delisle}, {Demouchy},
  {Dharmawardena}, {Diakite}, {Diener}, {Distefano}, {Dolding}, {Enke},
  {Fabre}, {Fabrizio}, {Fedorets}, {Fernique}, {Figueras}, {Fournier},
  {Fouron}, {Fragkoudi}, {Gai}, {Garcia-Gutierrez}, {Garcia-Reinaldos},
  {Garc{\'\i}a-Torres}, {Garofalo}, {Gavel}, {Gerlach}, {Geyer}, {Gilmore},
  {Girona}, {Giuffrida}, {Gomel}, {Gomez}, {Gonz{\'a}lez-N{\'u}{\~n}ez},
  {Gonz{\'a}lez-Santamar{\'\i}a}, {Gonz{\'a}lez-Vidal}, {Granvik}, {Guillout},
  {Guiraud}, {Guti{\'e}rrez-S{\'a}nchez}, {Guy}, {Hatzidimitriou}, {Hauser},
  {Haywood}, {Helmer}, {Helmi}, {Sarmiento}, {Hidalgo}, {H{\l}adczuk}, {Hobbs},
  {Holland}, {Huckle}, {Jardine}, {Jasniewicz}, {Jean-Antoine Piccolo},
  {Jim{\'e}nez-Arranz}, {Juaristi Campillo}, {Julbe}, {Karbevska}, {Khanna},
  {Kordopatis}, {Korn}, {K{\'o}sp{\'a}l}, {Kostrzewa-Rutkowska},
  {Kruszy{\'n}ska}, {Kun}, {Laizeau}, {Lambert}, {Lanza}, {Lasne}, {Le
  Campion}, {Lebreton}, {Lebzelter}, {Leccia}, {Lecoeur-Taibi}, {Liao},
  {Licata}, {Lindstr{\o}m}, {Lister}, {Livanou}, {Lobel}, {Lorca}, {Loup},
  {Madrero Pardo}, {Magdaleno Romeo}, {Managau}, {Mann}, {Manteiga},
  {Marchant}, {Marconi}, {Marcos}, {Marcos Santos}, {Mar{\'\i}n Pina},
  {Marinoni}, {Marocco}, {Marshall}, {Polo}, {Mart{\'\i}n-Fleitas}, {Marton},
  {Mary}, {Masip}, {Massari}, {Mastrobuono-Battisti}, {McMillan}, {Messina},
  {Michalik}, {Millar}, {Mints}, {Molina}, {Molinaro}, {Moln{\'a}r}, {Monari},
  {Mongui{\'o}}, {Montegriffo}, {Montero}, {Mor}, {Mora}, {Morbidelli},
  {Morris}, {Muraveva}, {Murphy}, {Musella}, {Nagy}, {Noval}, {Oca{\~n}a},
  {Ogden}, {Ordenovic}, {Osinde}, {Pagani}, {Pagano}, {Palaversa}, {Palicio},
  {Pallas-Quintela}, {Panahi}, {Payne-Wardenaar}, {Pe{\~n}alosa Esteller},
  {Penttil{\"a}}, {Pichon}, {Piersimoni}, {Pineau}, {Plachy}, {Plum}, {Poggio},
  {Pr{\v{s}}a}, {Pulone}, {Racero}, {Ragaini}, {Rainer}, {Raiteri}, {Ramos},
  {Ramos-Lerate}, {Regibo}, {Richards}, {Rios Diaz}, {Ripepi}, {Riva}, {Rix},
  {Rixon}, {Robichon}, {Robin}, {Robin}, {Roelens}, {Rogues}, {Rohrbasser},
  {Romero-G{\'o}mez}, {Rowell}, {Royer}, {Ruz Mieres}, {Rybicki}, {S{\'a}ez
  N{\'u}{\~n}ez}, {Sagrist{\`a} Sell{\'e}s}, {Salguero}, {Samaras}, {Sanchez
  Gimenez}, {Sanna}, {Santove{\~n}a}, {Sarasso}, {Schultheis}, {Sciacca},
  {Segol}, {Segovia}, {Semeux}, {Siddiqui}, {Siebert}, {Siltala}, {Silvelo},
  {Slezak}, {Slezak}, {Smart}, {Snaith}, {Solano}, {Solitro}, {Souami},
  {Souchay}, {Spagna}, {Spina}, {Spoto}, {Steele}, {Steidelm{\"u}ller},
  {Stephenson}, {S{\"u}veges}, {Surdej}, {Szabados}, {Szegedi-Elek}, {Taris},
  {Taylor}, {Teixeira}, {Tolomei}, {Tonello}, {Torra}, {Torra}, {Torralba
  Elipe}, {Trabucchi}, {Tsounis}, {Turon}, {Ulla}, {Unger}, {Vaillant}, {van
  Dillen}, {van Reeven}, {Vanel}, {Vecchiato}, {Viala}, {Vicente}, {Voutsinas},
  {Weiler}, {Wevers}, {Wyrzykowski}, {Yoldas}, {Yvard}, {Zhao}, {Zorec}, \&
  {Zucker}}]{Arenou2022}
{Gaia Collaboration}, {Arenou}, F., {Babusiaux}, C., {et~al.} 2022, arXiv
  e-prints, arXiv:2206.05595.
\newblock \doarXiv{2206.05595}

\bibitem[{{Goldin} \& {Makarov}(2007)}]{Goldin2007}
{Goldin}, A., \& {Makarov}, V.~V. 2007, \apjs, 173, 137, \dodoi{10.1086/520513}

\bibitem[{{Horch} {et~al.}(2017){Horch}, {Casetti-Dinescu}, {Camarata},
  {Bidarian}, {van Altena}, {Sherry}, {Everett}, {Howell}, {Ciardi}, {Henry},
  {Nusdeo}, \& {Winters}}]{Horch2017}
{Horch}, E.~P., {Casetti-Dinescu}, D.~I., {Camarata}, M.~A., {et~al.} 2017,
  \aj, 153, 212, \dodoi{10.3847/1538-3881/aa6749}

\bibitem[{{Luyten}(1979)}]{Luyten1979}
{Luyten}, W.~J. 1979, {NLTT catalogue. Volume\_I. +90\_\_to\_+30\_.
  Volume.\_II. +30\_\_to\_0\_.}

\bibitem[{{Mason} {et~al.}(2001){Mason}, {Wycoff}, {Hartkopf}, {Douglass}, \&
  {Worley}}]{WDS}
{Mason}, B.~D., {Wycoff}, G.~L., {Hartkopf}, W.~I., {Douglass}, G.~G., \&
  {Worley}, C.~E. 2001, \aj, 122, 3466, \dodoi{10.1086/323920}

\bibitem[{{Nidever} {et~al.}(2002){Nidever}, {Marcy}, {Butler}, {Fischer}, \&
  {Vogt}}]{Nidever2002}
{Nidever}, D.~L., {Marcy}, G.~W., {Butler}, R.~P., {Fischer}, D.~A., \& {Vogt},
  S.~S. 2002, \apjs, 141, 503, \dodoi{10.1086/340570}

\bibitem[{{Nordstr{\"o}m} {et~al.}(2004){Nordstr{\"o}m}, {Mayor}, {Andersen},
  {Holmberg}, {Pont}, {J{\o}rgensen}, {Olsen}, {Udry}, \& {Mowlavi}}]{N04}
{Nordstr{\"o}m}, B., {Mayor}, M., {Andersen}, J., {et~al.} 2004, \aap, 418,
  989, \dodoi{10.1051/0004-6361:20035959}

\bibitem[{{Paredes} {et~al.}(2021){Paredes}, {Henry}, {Quinn}, {Gies},
  {Hinojosa-Go{\~n}i}, {James}, {Jao}, \& {White}}]{Paredes2021}
{Paredes}, L.~A., {Henry}, T.~J., {Quinn}, S.~N., {et~al.} 2021, \aj, 162, 176,
  \dodoi{10.3847/1538-3881/ac082a}

\bibitem[{{Pecaut} \& {Mamajek}(2013)}]{Pecaut2013}
{Pecaut}, M.~J., \& {Mamajek}, E.~E. 2013, \apjs, 208, 9,
  \dodoi{10.1088/0067-0049/208/1/9}

\bibitem[{{Tokovinin}(2014)}]{FG67b}
{Tokovinin}, A. 2014, \aj, 147, 87, \dodoi{10.1088/0004-6256/147/4/87}

\bibitem[{{Tokovinin}(2015)}]{Tok2015}
---. 2015, \aj, 150, 177, \dodoi{10.1088/0004-6256/150/6/177}

\bibitem[{{Tokovinin}(2016{\natexlab{a}})}]{chiron1}
---. 2016{\natexlab{a}}, \aj, 152, 11, \dodoi{10.3847/0004-6256/152/1/11}

\bibitem[{{Tokovinin}(2016{\natexlab{b}})}]{orbit}
---. 2016{\natexlab{b}}, {Orbit: IDL Software For Visual, Spectroscopic, And
  Combined Orbits},  Zenodo, \dodoi{10.5281/zenodo.61119}

\bibitem[{{Tokovinin}(2017)}]{ORBIT3}
---. 2017, {ORBIT3: Orbits of Triple Stars},  Zenodo,
  \dodoi{10.5281/zenodo.321854}

\bibitem[{{Tokovinin}(2018{\natexlab{a}})}]{MSC}
---. 2018{\natexlab{a}}, \apjs, 235, 6, \dodoi{10.3847/1538-4365/aaa1a5}

\bibitem[{{Tokovinin}(2018{\natexlab{b}})}]{chiron4}
---. 2018{\natexlab{b}}, \aj, 156, 194, \dodoi{10.3847/1538-3881/aadfe6}

\bibitem[{{Tokovinin}(2019)}]{chiron6}
---. 2019, \aj, 158, 222, \dodoi{10.3847/1538-3881/ab4c94}

\bibitem[{{Tokovinin}(2020)}]{chiron7}
---. 2020, \aj, 160, 69, \dodoi{10.3847/1538-3881/ab9b1e}

\bibitem[{{Tokovinin}(2022)}]{chiron8}
---. 2022, \aj, 163, 161, \dodoi{10.3847/1538-3881/ac5330}

\bibitem[{{Tokovinin} {et~al.}(2013){Tokovinin}, {Fischer}, {Bonati},
  {Giguere}, {Moore}, {Schwab}, {Spronck}, \& {Szymkowiak}}]{CHIRON}
{Tokovinin}, A., {Fischer}, D.~A., {Bonati}, M., {et~al.} 2013, \pasp, 125,
  1336, \dodoi{10.1086/674012}

\bibitem[{{Tokovinin} {et~al.}(2010){Tokovinin}, {Hartung}, \&
  {Hayward}}]{Tok2010}
{Tokovinin}, A., {Hartung}, M., \& {Hayward}, T.~L. 2010, \aj, 140, 510,
  \dodoi{10.1088/0004-6256/140/2/510}

\bibitem[{{Tokovinin} {et~al.}(2012){Tokovinin}, {Hartung}, {Hayward}, \&
  {Makarov}}]{Tok2012}
{Tokovinin}, A., {Hartung}, M., {Hayward}, T.~L., \& {Makarov}, V.~V. 2012,
  \aj, 144, 7, \dodoi{10.1088/0004-6256/144/1/7}

\bibitem[{{Tokovinin} \& {Latham}(2017)}]{TL2017}
{Tokovinin}, A., \& {Latham}, D.~W. 2017, \apj, 838, 54,
  \dodoi{10.3847/1538-4357/aa6331}

\bibitem[{{Tokovinin} \& {L{\'e}pine}(2012)}]{TokLep2012}
{Tokovinin}, A., \& {L{\'e}pine}, S. 2012, \aj, 144, 102,
  \dodoi{10.1088/0004-6256/144/4/102}

\bibitem[{{Tokovinin} {et~al.}(2022){Tokovinin}, {Mason}, {Mendez}, \&
  {Costa}}]{Tokovinin2022}
{Tokovinin}, A., {Mason}, B.~D., {Mendez}, R.~A., \& {Costa}, E. 2022, \aj,
  164, 58, \dodoi{10.3847/1538-3881/ac78e7}

\bibitem[{{Tokovinin} {et~al.}(2020){Tokovinin}, {Mason}, {Mendez}, {Costa}, \&
  {Horch}}]{Tok2020}
{Tokovinin}, A., {Mason}, B.~D., {Mendez}, R.~A., {Costa}, E., \& {Horch},
  E.~P. 2020, \aj, 160, 7, \dodoi{10.3847/1538-3881/ab91c1}

\bibitem[{{Tokovinin} {et~al.}(2015){Tokovinin}, {Pribulla}, \&
  {Fischer}}]{LCO}
{Tokovinin}, A., {Pribulla}, T., \& {Fischer}, D. 2015, \aj, 149, 8,
  \dodoi{10.1088/0004-6256/149/1/8}

\bibitem[{{Tokovinin} \& {Smekhov}(2002)}]{TS2002}
{Tokovinin}, A.~A., \& {Smekhov}, M.~G. 2002, \aap, 382, 118,
  \dodoi{10.1051/0004-6361:20011586}

\end{thebibliography}

\end{document}